\documentclass{pasj00}

\begin{document}
\SetRunningHead{Author(s) in page-head}{Running Head}
\Received{2008/05/23}
\Accepted{2008/11/06}

\title{Deep 15 $\mu m$ AKARI observations in the CDFS : estimating dust luminosities for a MIR-selected sample and for Lyman Break Galaxies and the evolution of $L_{dust} / L_{UV}$ with the redshift.}

\author{Denis \textsc{Burgarella} %
  \thanks{Example: Present Address is xxxxxxxxxx}}
\affil{Laboratoire d'Astrophysique de Marseille, Observatoire Astronomique Marseille Provence, Universit\'e d'Aix-Marseille, CNRS}
\email{denis.burgarella@oamp.fr}

\author{V\'eronique \textsc{Buat}}
\affil{Laboratoire d'Astrophysique de Marseille, Observatoire Astronomique Marseille Provence, Universit\'e d'Aix-Marseille, CNRS}\email{veronique.buat@oamp.fr}
\author{Tsutomu T. {\sc Takeuchi}}
\affil{Institute for Advanced Research, Nagoya University, Furo-cho, Chikusa-ku, Nagoya 464--8601, Japan}\email{takeuchi@iar.nagoya-u.ac.jp}
\author{Takehiko {\sc Wada}}
\affil{Department of Infrared Astronomy, ISAS, JAXA}\email{wada@ir.isas.jaxa.jp}
\and
\author{Chris {\sc Pearson}}
\affil{Space Science and Technology Department, CCLRC Rutherford Appleton
Laboratory, Oxford}
\affil{Department of Physics, University of Lethbridge, 4401 University Drive,
Lethbridge, Alberta T1J 1B1, Canada}\email{c.pearson@rl.ac.uk}

%

\KeyWords{galaxies: formation - galaxies: evolution - galaxies: high redshift - infrared: galaxies - cosmology: observations} 

\maketitle

\begin{abstract}
Deep observations of the Chandra Deep Field South have been secured at $15 \mu m$ with AKARI/IRC infrared space telescope. From these observations, we define a sample of mid infrared-selected galaxies at $15 \mu m$ and we also obtain $15 \mu m$ flux densities for a sample of Lyman Break Galaxies at $z \sim 1$ already observed at $24 \mu m$ with Spitzer/MIPS. Number counts for the mid infrared-selected sample show a bump around a $15 \mu m$ flux density of $0.2 mJy$ that can be attributed to galaxies at $z > 0.4$ and at $z > 0.8$ for the fainter part of the bump. This bump seems to be shifted as compared to other works and a possible origin can be the Cosmic variance. Thanks to this dataset, we have tested, on the two above samples $z \sim 1$, the validity of the conversions from monochromatic luminosities $\nu  f_{\nu}$ at a rest-frame wavelength of $8 \mu m$ by a comparison with total dust luminosities estimated from Spitzer rest-frame $12 \mu m$ data that we use as a reference. We find that the $8 \mu m$ dust luminosities are not all consistent and that some of them are better when compared to $L_{dust}$ evaluated from longer wavelength luminosities. We also find that the rest-frame $8\mu m$ luminosities provide globally good estimates of $L_{dust}$.  By comparing our data for the two samples to several libraries of spectral energy distributions, we find that models can explain the diversity of the observed $f_{24} / f_{15}$ ratio quite reasonably for the MIR-selected sample and better for the LBG sample which are less dispersed than the MIR selection. However, when we analyse the luminosity dependence of this ratio, we find important discrepancies. Finally, we revisit the evolution of $L_{dust} / L_{UV}$ ratio with the redshift $z$ by re-calibrating previous $L_{dust}$ at $z \sim 2$ based on our results and added new data points at higher redshifts. The decreasing trend is amplified as compared to the previous estimate. 
\end{abstract}

\section{Introduction}

The star formation rate (SFR) of galaxies can be seen as some kind of Sesame that opens up the door of a quantitative understanding of the formation and the evolution of galaxies at all redshifts. It is one of the main tools of modern extragalactic astrophysics and astronomers world-wide have been looking for the best one(s) for several decades.

All young stars in galaxies emit ultraviolet (UV) photons that can be collected by telescopes to estimate the SFR of these galaxies (e.g. \cite{leitherer99}). However young stars might be embedded in dust that prevents part of the UV photons to reach our telescopes. These UV photons are scattered and absorbed by dust grains and reprocessed into infrared (IR) photons (e.g. \cite{buat96}) which poses a problem to obtain a reliable SFR than can be seen as a total one. 

Several methods have been proposed to evaluate the total amount of SFR, including that part hidden in dust which cannot be measured by UV/optical telescopes.  \citet{meurer99} proposed a version based on a correction for the dust attenuation that would be applied to the UV flux. This correction relies only on UV data and is estimated from the UV slope $\beta$ (assuming $f_\lambda \propto \lambda^{\beta}$): the redder the UV slope the larger the correction. However, this method seems not to be fully reliable for any type of galaxies (\cite{buat05}) or, at least, provides an estimate that is not very accurate. Usually, the calibration is performed from infrared far-IR (FIR) data  but this happens not to be that simple because of various technological challenges: the low transparency of the atmosphere that implies to launch space telescopesl to observe above a few micrometers, the necessity to develop efficient detectors in this wavelength range and, of course, the mandatory cooling of these telescopes. Moving from the FIR to the mid-IR (MIR) looks like a good idea since the MIR emission is easier to measure but, again, there might be issues that need to be carefully checked. First, the MIR range can be subject to a contamination from stellar emission (especially when observing high redshift objects). Next, some polycyclic aromatic hydrocarbons (PAHs) emission features are present in this range that are correlated to the metallicity (\cite{engelbracht08}). Then, most of the SFR relationships have been calibrated from (or at least mostly rely on) observations of local galaxies. We can wonder whether they are also valid at higher redshifts. Indeed, the spectral energy distributions (SEDs) of galaxies can evolve with the redshift and we do not definitely know, for instance, whether or not low and high redshifts share the same PAH emissions in the MIR.

Moreover, even when higher redshift objects have been accounted for (e.g. \cite{bavouzet08}), these galaxies are not UV-selected galaxies. So, we have to check whether the relationships developped from IR samples are valid for UV-selected samples.
 
Our objective, in this paper, is to use two samples of galaxies: one selected from deep $15 \mu m$ observations of the Chandra Deep Field South (CDFS) and one sample of Lyman Break Galaxies (LBGs) at $z \sim 1$ from GALEX (\cite{burgarella06} and \cite{burgarella07}) to compare estimates of the dust luminosities (computed from rest-frame $8 \mu m$ -- i.e. AKARI observations at $15 \mu m$ -- using formulae provided by several authors) to the ones evaluated from Spitzer observed flux densities at $24 \mu m$ (i.e. rest-frame $12 \mu m$ at $z = 1$). Then, we will test the MIR SEDs through a comparison of the observed  $f_{24} / f_{15}$ flux ratio with several libraries of SEDs. 

Recent works by \citet{reddy06} and \citet{burgarella07} suggest that the dust-to-UV luminosity ratio ($L_{dust} / L_{UV}$) would decrease from $z = 0$ to $z = 2$ for a galaxy with a given luminosity $L_{dust}$. However, this effect is based on a comparison of datasets at different redshifts and we need to revisit this effect after checking the conversions from MIR luminosites to $L_{dust}$.

In Sect. 2, we will present the observations of the Chandra Deep Field South with AKARI/IRC camera through the L15 filter centered at $15 \mu m$. Sect. 3 will present the two samples that we will use in this paper. Then, we will test several calibrations of $8 \mu m$ MIR luminosities to total dust luminosities.  In Sect. 4, we will study the evolution of IR-to-UV luminosity ratio with the redshift by accounting for the results found in the paper. Finally, Sect. 5 will present the spectral energy distribution libraries that we will use. We will try to understand better the physics of the two galaxy samples by analysing the $f_{24} / f_{15}$ ratio by a comparison with the models and templates. 

We assume a cosmology with $H_0 = 70 ~km.s^{-1}.Mpc^{-1}$, $\Omega_M = 0.3$ and $\Omega_{\Lambda}=0.7$.

\section{Observations and Data Reduction}
The AKARI satellite  \citep{murakami07} is a Japanese space mission dedicated to infrared astrophysics and was launched on board JAXA's M-V8 Launch Vehicle on February 22, 2006 Japan Standard Time, JST. AKARI has a 68.5 cm cooled telescope with two focal plane instruments, the Far-Infrared Surveyor (FIS) \citep{kawada07} and the Infrared Camera (IRC).  The IRC consists of three cameras, the IRC-NIR, MIR-S \& MIR-L covering 1.7-- 26 $\mu$m in 9 bands with fields of view (FoV) of 10$\arcmin$  $\times$ 10$\arcmin$ \citep{onaka07}. Although the NIR/MIR-S cameras share the same field of view, the MIR-L camera is separated on the sky by 20 arcmin. AKARI observations are segregated into Large Surveys (LS) and guaranteed time Mission Programs. In addition, 10$\%$ of the total observation time was reserved for Open Time observations from ESA member institutes. In this work, we report on our Open Time program (IRLBG, P.I. D. Burgarella) to image the Chandra Deep Field South (CDFS: RA=53.0465, Dec=-27.8307, J2000) in the AKARI-IRC $15 \mu$m band (the L15 band covering the wavelength range 12.6 -- 19.4$\mu$m). A total of 5 pointed  observations were taken on the 31st January and 1st February 2007, using the  IRC Astronomical Observing Template (AOT) designed for deep observations (IRC05)  in the configuration $b;L$, corresponding to the  L15 filter on the IRC-MIR-L camera (See Table \ref{tab:obslog}). In this configuration, each pointing consists of approximately 10 exposure cycles. Each exposure cycle consists of a single short (0.6s) exposure frame and three long (19s) exposure frames. For deep imaging, the short exposure frames are insignificant and are often discarded. Note that, for a target centered on the FoV of the MIR-L, data is also taken in the NIR/MIR-S cameras, albeit offset by $\sim$20 arcmin with the FoV rotating around the MIR-L target position.  The IRC05 AOT has no explicit dithering built into the AOT, therefore dithering is achieved by carrying out at least three pointed observations with this AOT at any given position. The dithering is achieved from the natural slight offset movement and slight rotation of the field of view between any pair of pointed observations

The data were reduced by the standard IRC imaging pipeline (version 070911; see IRC Data User Manual by \cite{lorente-pipeline}). The pipeline corrects basic instrumental effects, performing dark subtraction, linearity correction, distortion correction and flat fielding. The superdark frame version 061109  and super flat version 070912 were used. In order to reduce the effect from the numerous hot pixels in the MIR-L detectors, hot pixels were identified and set to a very large numeric value such that they would be identified and removed as cosmic ray events in a later processing stage of the pipeline. The pixel values were then interpolated using the surrounding pixel values after the flat fielding. 

AKARI observations suffer from the effects of stray light from Earthshine and we have rejected any frames whose sky value had a measured standard deviation significantly larger than the mean between frames. Before combining the image from individual exposure frames for a single pointing, the  frames have a median sky subtracted. Alignment of the frames for co-addition is usually determined by the position of bright point sources in each frame. However, in case of the MIR-L channel, the number of the point sources which had sufficient signal-to-noise were too few. Therefore, in order to align the MIR-L frames, we identified common objects in our MIR-L frame and in the Spitzer $24 \mu m$ FIDEL image (http://ssc.spitzer.caltech.edu/legacy/).

\begin{figure*}
  \begin{center}
   \FigureFile(15cm,15cm){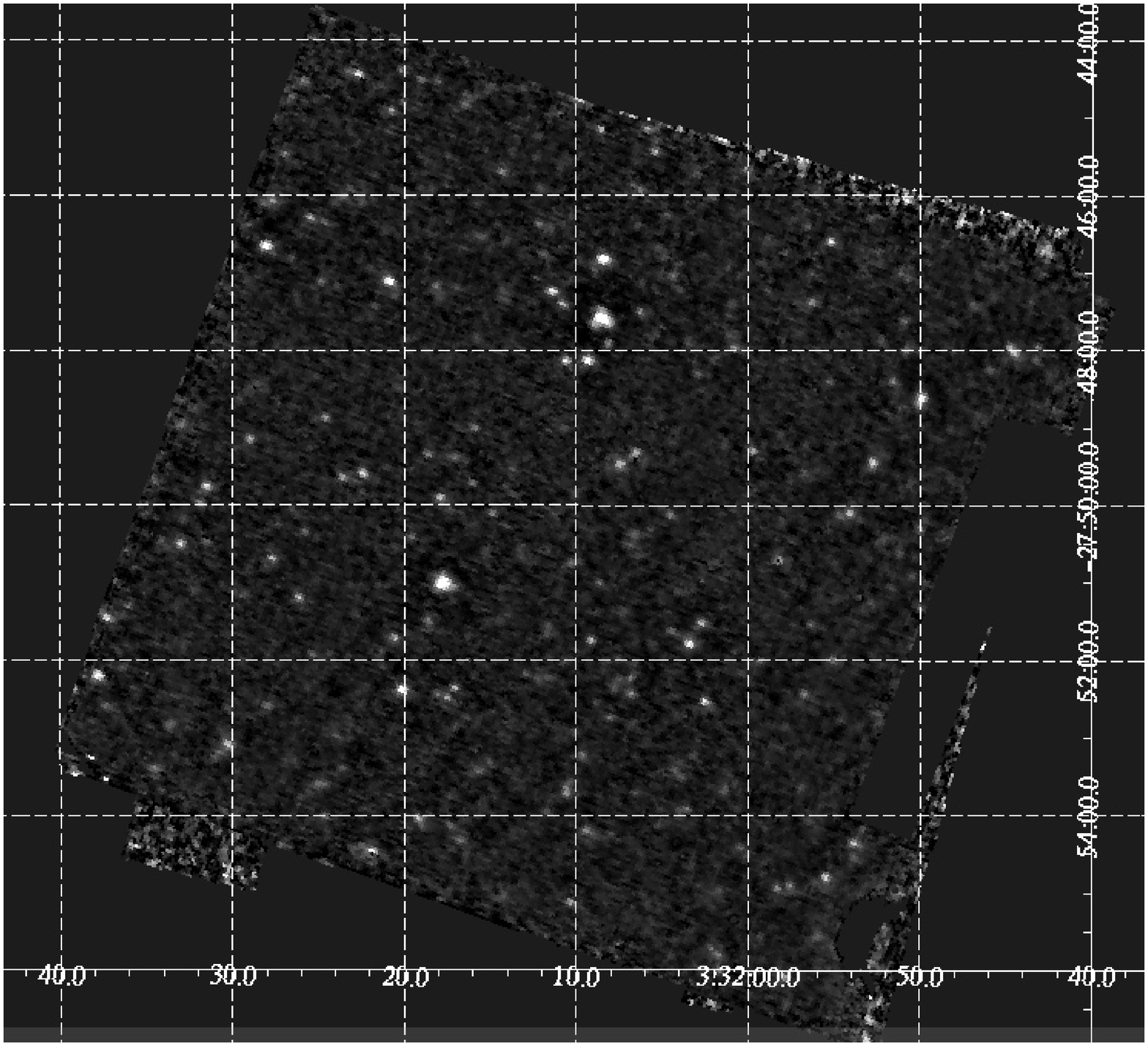}
  \end{center}
  \caption{Combined AKARI / IRC 15 $\mu m$ frame}\label{15umframe}
\end{figure*}

Spurious events, such as cosmic rays and electronic ghost images are removed at the co-addition stage using a 3 sigma clipping technique. For our data, given the large number of frames available for co-addition, we used the ccdclip option in the pipeline, where the detector noise performance is assumed as opposed to the default option sigclip where the noise performance is estimated from the data itself. The average rather than the median  value among the frames, for each pixel was used in order to further improve the signal-to-noise of the final image. For every pointing a total of 30 exposure frames were co-added corresponding to a total exposure time of 570s. The five individual pointings were then combined to make the final image (Fig. \ref{15umframe}).

\begin{table*}
  \caption{Observation Log}\label{tab:obslog}
  \begin{center}
    \begin{tabular}{lllll}
      \hline
      Target id & AOT & OBS-DATE &  RA & DEC \\
4040001-001 & IRC05 b;L & 2007-01-31 04:05:30 & 53.0465 & -27.8307 \\
4040001-002 & IRC05 b;L & 2007-01-31 05:44:49 & 53.0465 & -27.8307 \\
4040001-003 & IRC05 b;L & 2007-01-31 23:57:15 & 53.0465 & -27.8307 \\
4040001-004 & IRC05 b;L & 2007-02-01 01:36:34 & 53.0465 & -27.8307 \\
4040001-005 & IRC05 b;L & 2007-02-01 03:15:52 & 53.0465 & -27.8307 \\
      \hline
    \end{tabular}
  \end{center}
\end{table*}

\section{Definition of the MIR-selected and the Lyman Break Galaxy samples}

The above AKARI image has been used to build two galaxy samples : a $15 \mu m$ - selected (or equivalently MIR-selected) sample defined from the photometry carried out on the image. We also measured the $15 \mu m$ flux density of a sample of Lyman Break Galaxy (LBG) already defined from GALEX far ultraviolet (FUV) and near ultraviolet(NUV) images (\cite{burgarella07}).

\subsection{The MIR - selected sample}

The IRAF task DAOFIND has been used to identify a set of objects at a $3-\sigma$ level above the background. To minimise confusion errors, we have integrated over an aperture of 2 pixels using the IRAF task PHOT and then, this value has been extrapolated (by a factor of 1.645 estimated from a PSF built from several point sources in the field) to an integrated flux density over 7.5 pixels to match the recommendation of the AKARI / IRC manual by \citet{lorente-pipeline}. Then, the calibration provided by (\cite{lorente-pipeline}, i.e. $f_{\nu} = f_{7.5pixels} \times 1.48$) has been applied to the data to get flux densities in $\mu Jy$. Uncertainties on the flux densities are provided by PHOT. The estimate of the uncertainties also account for the 2.82 \% error on the calibration constant quoted by \citet{lorente-pipeline}. A list of 323 MIR - selected objects is created. Then, we have cross-correlated this sample with COMBO 17 (\cite {wolf04}). Whenever several counterparts are found within the 2-arcsec radius, we select the best one by eye. The final sample that we use in this paper contains 239 objects that we also have cross-correlated with the $24 \mu m$-selected sample from \cite{buat08}. The dataset is presented Table \ref{tab:15umsample}.

\begin{figure*}
  \begin{center}
   \FigureFile(15cm,15cm){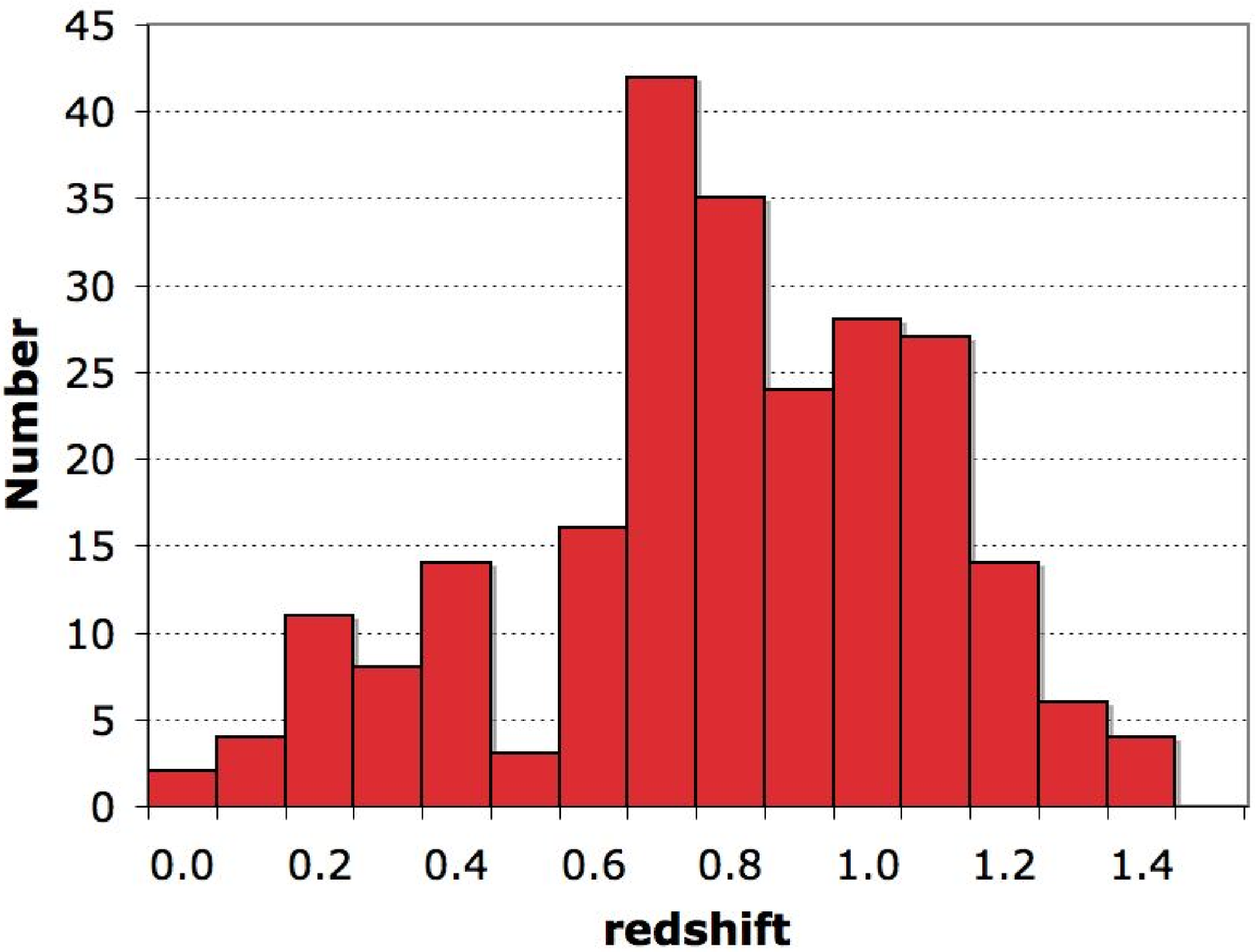}
  \end{center}
  \caption{The redshift distribution is evaluated by using COMBO 17 redshifts.}\label{zdistr}
\end{figure*}

The redshift distribution of this sample is shown in Fig. \ref{zdistr}. It shows that galaxies up to $z = 1.3$ have been detected in the AKARI image. We select only objects at $z \le 1.2$ for this analysis because the quality of the highest redshift objects is too low.

\begin{table*}
  \caption{$15\mu m$-selected sample}\label{tab:15umsample}
  \begin{center}
    \begin{tabular}{lllllllll}
      \hline
      COMBO 17 Id & RA	 & Dec &  z & $s_{15\mu m}$ & $errs_{15\mu m}$ & $s_{24\mu m}$ & $errs_{24\mu m}$ & $L_{dust}$ \\
--- & --- & --- & --- & --- & --- & --- & --- & --- \\
      \hline
    \end{tabular}
  \end{center}
\end{table*}

\subsection{The $z \sim 1$ Lyman Break Galaxy sample}

\citet{burgarella07} defined a sample of 420 LBGs at $0.9 < z < 1.3$ from ultraviolet images in the deep FUV and NUV exposures in the CDFS. In this paper, the authors find that this LBG sample shares its main characteristics with UV-selected samples. The LBG sample is formed of 80\% not detected at $24\mu m$ (we called them Blue LBG = BLBGs) and 20\% detected at $24 \mu m$ (Red LBGs = RLBGs). Note that this classification is only representative of a difference in dust attenuation but the two of them share the same other properties in average (star formation history, morphology, etc.). However, those 20\% of RLBGs contain 80\% of the SFR density of LBGs. \citet{burgarella07} also claim that the UV - optical spectral energy distributions (SEDs) of the $z \sim 1$ sample is very simular to the $z \sim 3$ LBG sample selected by \citet{forster04}. 

One of the initial motivation of the AKARI program is to check whether we could use the rest-frame $\sim 8 \mu m$ flux density to estimate the hidden SFR of LBGs (or UV-selected objects) at high redshift. Several authors have calibrated this rest-frame $\sim 8 \mu m$ flux density into a hidden star formation rate by using $z = 0$ galaxies (e.g. \cite{chary01}, \cite{takeuchi05b}, \cite{reddy06}). 

We have used the coordinates from the $z \sim 1$ LBG sample (\cite{burgarella07}) as inputs for the IRAF task phot, in the same way as for the MIR-selected sample. After cleaning up the results from the photometry, 27 LBGs are found to have a $15\mu m$ flux density within the range $17.39 \le s_{15\mu m}~[\mu Jy] \le 227.78$. Out of these 28 LBGs, 9 have been detected at $24 \mu m$ in the redshift range $0.9 \le z \le 1.2$ used in the rest of the paper. This small LBG dataset only contain RLBGs which are IR-bright objects. Moreover, since these galaxies are LBGs, they have been detected in the GALEX near-UV bands. This insures that we are not using higher redshift galaxies as $z \sim 1$ ones : at $z \sim 1.4$, the Lyman break would be beyond the red limit of the near-UV filter and no flux at all should be detected in this filter. The dataset is presented in Table \ref{tab:lbgsample}.

\begin{table*}
  \caption{$z \sim 1$ Lyman Break Galaxy sample}\label{tab:lbgsample}
  \begin{center}
    \begin{tabular}{lllllllll}
      \hline
      COMBO 17 Id & RA	 & Dec &  z & $s_{15\mu m}$ & $errs_{15\mu m}$ & $s_{24\mu m}$ & $errs_{24\mu m}$  & $L_{dust}$\\
--- & --- & --- & --- & --- & --- & --- & --- & --- \\
      \hline
    \end{tabular}
  \end{center}
\end{table*}

\section{The $15 \mu m$ number counts}

\begin{figure*}
  \begin{center}
    \FigureFile(10cm,10cm){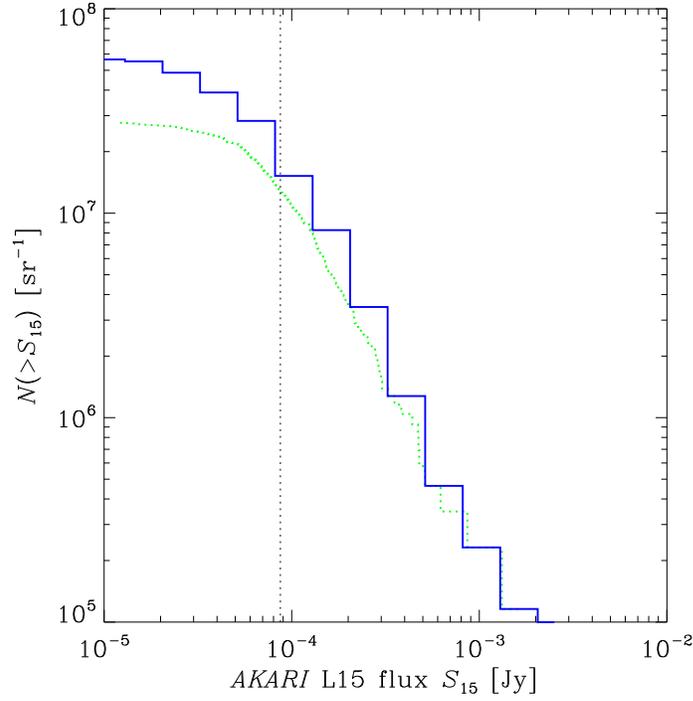}
  \end{center}
  \caption{The integral number counts (NC) of galaxies in the CDFS observed with AKARI IRC MIR L15 band. The green irregular step function (dotted line) shows the raw NC of the galaxies, while the regular step function (solid line) represents the NC corrected for the completeness. The vertical dotted line show the 80-\% flux completeness limit, which is found to be 85~$\mu$Jy (see text).}\label{fig:nc}
\end{figure*}

In this section, we show the number counts of AKARI 15-$\mu$m sample in the CDFS. The total number of the 15-$\mu$m galaxies in this field is 239.
Cumulative number counts (NC) of these galaxies are shown in Fig.~\ref{fig:nc} (dotted irregular step function). The source density is converted to the units of $[\mbox{sr}^{-1}]$ in Fig.~\ref{fig:nc}.

\begin{figure*}
  \begin{center}
    \FigureFile(10cm,10cm){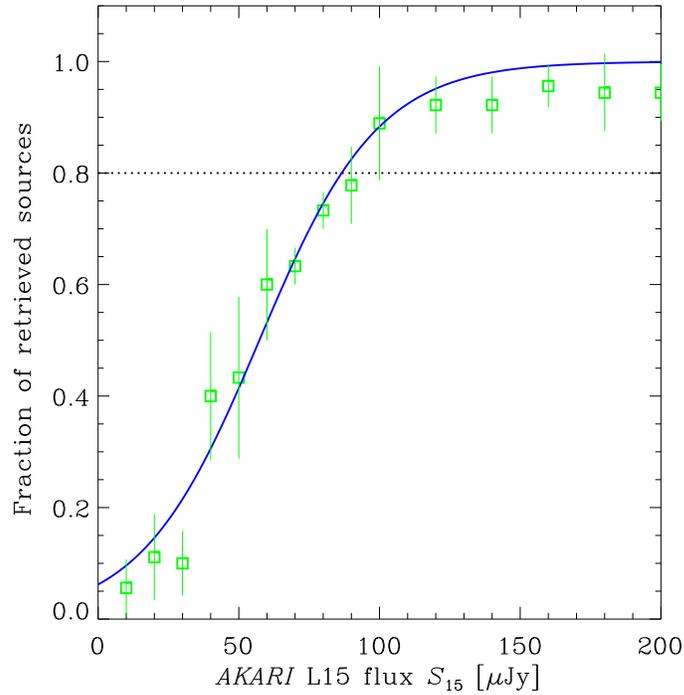}
  \end{center}
  \caption{Completeness of the source extraction of our CDFS   AKARI 15-$\mu$m image.   The open squares with error bars are the measures rate of retrieving 
  artificial sources, and the solid curve shows the fitted function.  We found that the 80-\% flux completeness limit is $85\;\mu$Jy.
}\label{fig:completeness}
\end{figure*}

We can see the effect of a possible incompleteness below a flux density of $\sim 100\;\mu$Jy. We have examined the incompleteness of the source extraction by adding additional sources with various flux densities on the original observed AKARI 15-$\mu$m image (IRAF task ADDSTAR) and running the source extracting task (IRAF DAOFIND). We can estimate the incompleteness by measuring how much fraction of the added artificial sources are detected by the source extraction procedure. We show the result of this examination in Fig.~\ref{fig:completeness}. The open squares with error bars are the measures rate of retrieving artificial sources. To have a continuous function of the completeness or ``missing rate'', we fit the measured value with a function of the Fermi-Dirac form,
\begin{equation}
  f(S_{15}) = \frac{1}{\exp \left(-\frac{S_{15} - S_{\rm C}}{\phi} \right)+1}
\label{eq:fd}
\end{equation}
where we found that $S_{\rm C} = 57.3$ and $\phi=21.1$.
By the intersection of this function and 0.8 (horizontal line in Fig.~\ref{fig:completeness}), we found that the 80-\% flux completeness limit is $85\;\mu$Jy. 
With eq.~\ref{eq:fd}, we can correct the raw NC and obtain the completeness-corrected NC. The blue regular step function in Fig.~\ref{fig:nc} is the corrected NC.

\begin{figure*}
  \begin{center}
    \FigureFile(10cm,10cm){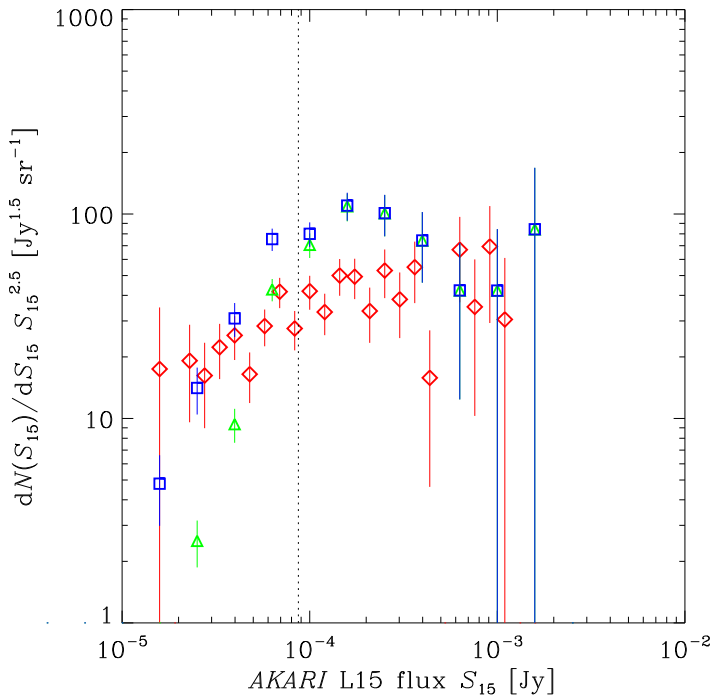}
  \end{center}
  \caption{Differential NC of the CDFS AKARI 15-$\mu$m galaxies. The green open triangles represent the raw differential NC, while the blue open squares show the differential NC corrected for the incompleteness. Error bars are calculated assuming Poissonian fluctuation. The red open diamonds are from \citet{wada07} in the NEP.
  }\label{fig:dc}
\end{figure*}

Based on the integrated NC, we obtain the differential NC in a standard manner. We show the Euclidean-normalized differential NC in Fig.~\ref{fig:dc}, i.e., $dN/dS_{15}$ multiplied by ${S_{15}}^{2.5}$. The open triangles represent the raw differential NC from our CDFS dataset and red diamonds represent results from \citet{wada07} in the NEP.  The errors are Poissonian. Again, with eq.~\ref{eq:fd}, we can correct the incompleteness effect. The open squares in Fig.~\ref{fig:dc} show the corrected differential NC. Error bars are calculated assuming Poissonian fluctuations in the original counts, and multiplied the incompleteness for those with corrected counts. The vertical dotted line is the same as Fig.~\ref{fig:nc}. As we can see, both differential NC are consistent but show a variation around the peak of CDFS observations and the AKARI Performance Verification (PV) one. However,
the cosmic variance in the CDFS (GOODS, to be precise) is estimated to be at least 60 \% in density-related measurements  \citep{somerville04}. Since the area is smaller than the whole GOODS area, we
should have larger variance both in CDFS $15 \mu m$ and in the PV field. In addtiion, is known that the CDFS hosts clusters at $z \sim 0.7$ (e.g.  \citet{adami05}). So, the observed discrepancy would not be statistically significant with respect to the cosmic variance (especially at $15 \mu m$ i.e. $7 \mu m$ rest-frame at $z \sim 1$ which is sensitive to any star formation in the clusters) and to other datasets presented in Figures 12 and 13 from \citet{wada07}. Quite surely we are observing the same population of galaxies as they do.

\begin{figure*}
  \begin{center}
    \FigureFile(10cm,10cm){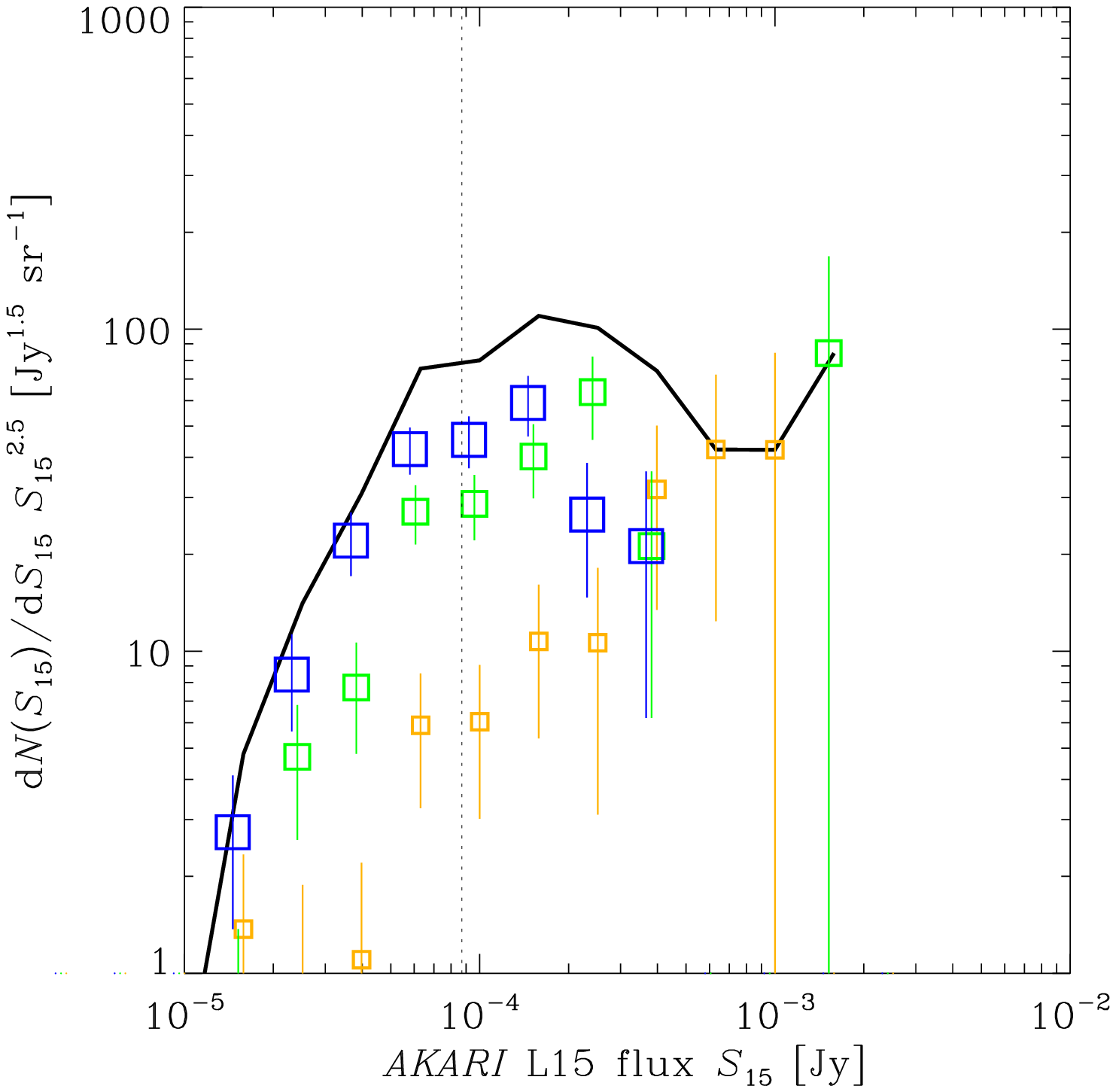}
  \end{center}
  \caption{Differential NC of the CDFS AKARI 15-$\mu$m galaxies, subdivided into galaxies with three redshift bins (completeness corrected). The smallest open squares represent the galaxies at $0 < z < 0.4$, the intermediate ones are at $0.4 < z < 0.8$, and the largest ones are $0.8 < z$. The broken solid line is the total differential NC, the same as shown in Fig.~\ref{fig:dc}. Clearly, the bump is dominated by galaxies with $0.4 < z$.
  }\label{fig:zdc}
\end{figure*}

The most prominent feature of the differential NC is the bump seen around a flux density of $200\;\mu$Jy. In order to see what population consists of this bump, we divide the differential counts into galaxies with three redshfit bins: $0 < z < 0.4$, $0.4 < z < 0.8$, and $0.8 < z$. We show the subdivided differential counts in Fig.~\ref{fig:zdc}. The smallest open squares, intermediate ones, and the largest ones represent galaxies at $0 < z < 0.4$, $0.4 < z < 0.8$, and $z > 0.8$, respectively. The broken solid line is the total differential NC, the same as shown in Fig.~\ref{fig:dc}. It is clear that the bump is dominated by galaxies with $z > 0.4$. Especially, the fainter part of the bump consists of galaxies at $z > 0.8$. This may mean that galaxy population observed at $15\;\mu$m experienced a significant evolution, as suggested by previous studies based on number counts \citep{pearson96,pearson05,takeuchi01,xu01}.

\section{SED Libraries}

\subsection{Spectra, templates and models in the Mid Infrared}

We have used several libraries of templates to be compared to the data. First, three  SED libraries commonly used to analyse MIR properties of local and distant galaxies: those of \citet{chary01} (CE),  \citet{dale02} (DH) and  \citet{lagache04} (LDP). These differents sets have been developped and optimized for different purposes: the CE library reproduces the local correlations between different luminosities from 7 to 850 $\mu$m, the DH templates describe the infrared SEDs of local normal star forming galaxies as a function of a single parameter, $f_{60}/f_{100}$, and the LDP templates of starburst and normal galaxies are built and optimized to reproduce statistical quantities like number counts, redshift distributions and  the cosmic infrared background. The original templates for starburst galaxies \citep{lagache03} have been modified in the MIR to reproduce the new Spitzer number counts with a decrease of the amplitude of the 11 $\mu$m feature by a factor $\sim 2$ \citep{lagache04}. The LDP templates for normal galaxies all exhibit the same shape and are only scaled in luminosity. They are similar to DH templates with intermediate $f_{60}/f_{100}$ values and thus, moderate luminosities. So, we do not reproduce all of them here.

We have also considered recent templates built from the low-resolution Spitzer IRS spectra. From their analysis of IRS spectra for 59 nearby galaxies \citet{smith07} have proposed five templates from 5 to 38 $\mu$m, one is a luminosity weighted average of all spectra, the other ones represent averages of spectra with different strengths of the features inside the PAH bands (weak AGNs are also included in the sample). To enlarge the range of observed spectra we have also included in our set of templates the nine classes of spectra proposed by \citet{spoon07}. Their analysis is based on the observation of a sample of AGNs, starburst galaxies together with LIRGs and ULIRGs. Their MIR templates exhibit very different spectral shapes from  continuum dominated spectra for AGNs to  spectra dominated by the 10 $\mu$m silicate trough and PAH-dominated starburst spectra. Last but not least, we have also included the starburst template of \citet{brandl06} obtained by averaging the IRS spectra of 13 nearby starburst galaxies.

\begin{figure*}
  \begin{center}
   \FigureFile(18cm,18cm){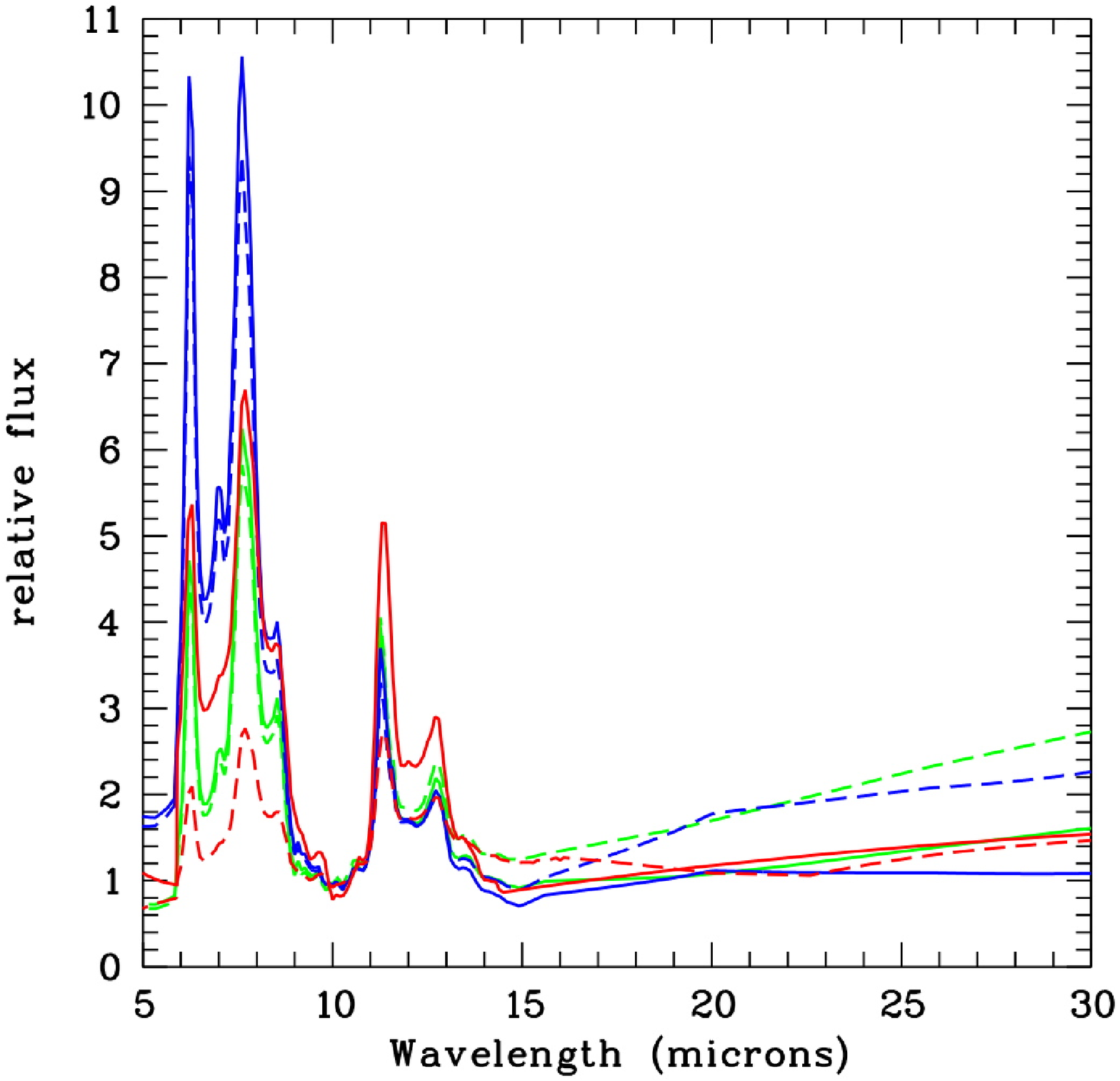}
   \end{center}
  \caption{Sets of templates at $L_{dust}=10^{11}$ L$\odot$ (solid lines) and $L_{dust}=10^{12}$ L$\odot$ (dashed lines)) for the libraries of \citet{chary01} (red), \citet{dale02} (green) and \citet{lagache04} (blue),all the spectra are normalized at 10 $\mu$m, see text for details}\label{CEtemplates}
\end{figure*}

\begin{figure*}
  \begin{center}
   \FigureFile(15cm,15cm){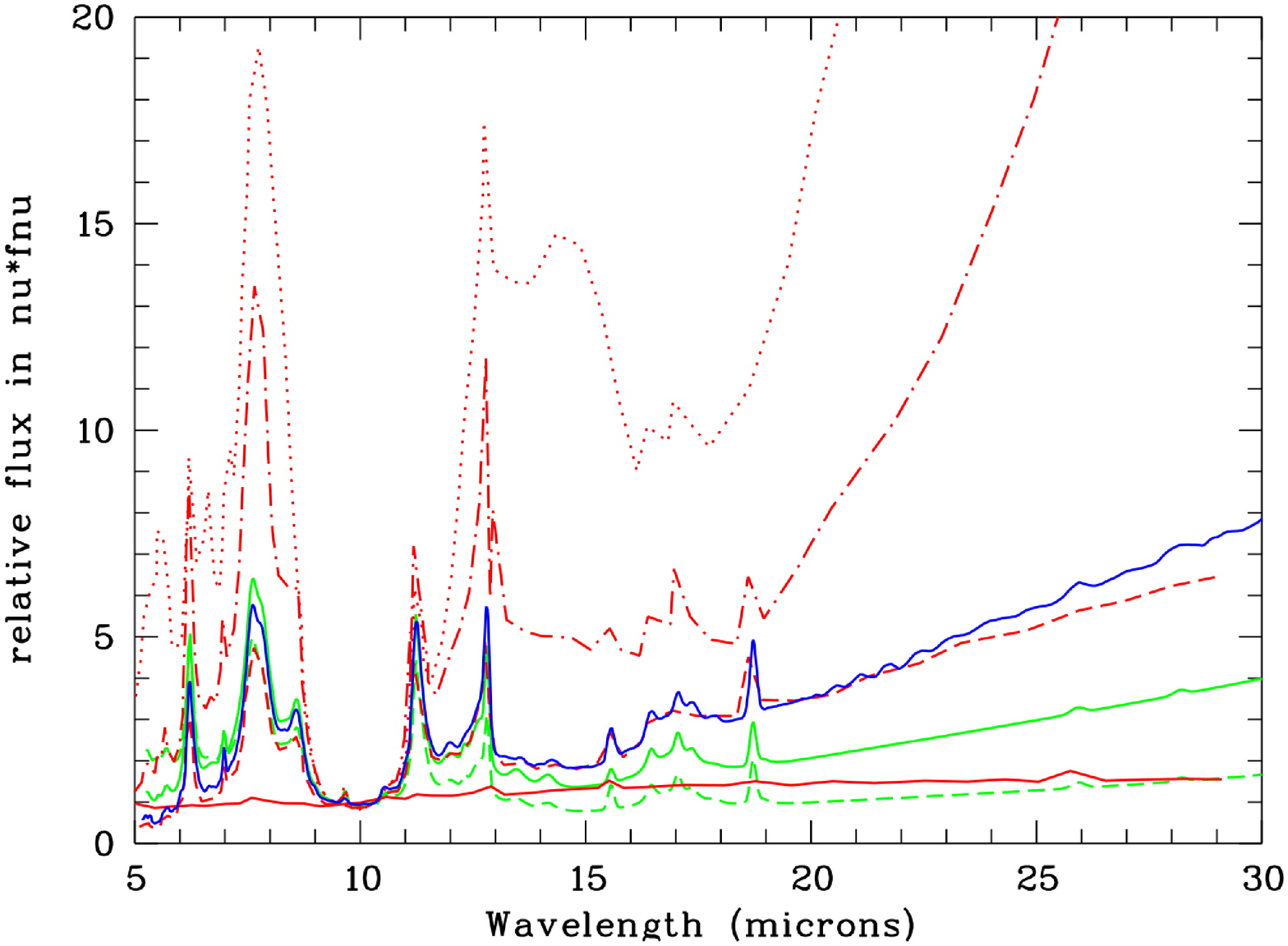}
   \end{center}
\caption{A few templates based on Spitzer/IRS spectra. The red lines corresponds to the templates of \citet{spoon07} 1A (solid line), 3B (dotted line), 1C (dashed line) and 2C (dot dashed line). The green lines refer to \citet{smith07} templates: the luminosity averaged template (solid line) and the template 3 (dashed line). The starburst template of \citet{brandl06} is represented by the blue solid line. All the spectra are normalized at 10 $\mu$m,  see text for details on the choice of the templates}\label{IRStemplates}
\end{figure*}

The different sets of templates are illustrated in Fig. \ref{CEtemplates} and Fig. \ref{IRStemplates} where a few representative spectra for each library are plotted. For the CE and LDP libraries (Fig. \ref{CEtemplates}) we have chosen the templates corresponding to galaxies with  $L_{dust}$ equal to $10^{11}$ and $10^{12}$ L$\odot$. The DH templates are given as a function of $f_{60}/f_{100}$. For the purpose of our comparison between the differents libraries, it is  convenient to attribute to them a total dust luminosity. At this aim, we follow \citet{marcillac06} who fitted a linear relation between the logarithms of $L_{dust}$ and $f_{60}/f_{100}$ valid for the nearby universe. We have plotted the templates with  log($f_{60}/f_{100}$)=-0.2 and log($f_{60}/f_{100}$)=-0.1 which correspond respectively to a total dust luminosity of the order of $10^{11}$ and $10^{12}$ L$\odot$. The templates obtained from Spitzer/IRS spectra are plotted in Fig. \ref{IRStemplates}: the luminosity-weighted average spectrum and one other spectrum (template 3) from  \citet{smith07}, the templates 1A (continuum dominated), 3B (prominent silicate absorption), 1C and 2C (PAH dominated) from \citet{spoon07} and the starburst template of \citet{brandl06}.

\subsection{The rest-frame $8 \mu m$ emission redshifted at $z = 1$ and $z = 2$}

\begin{figure*}
  \begin{center}
   \FigureFile(15cm,15cm){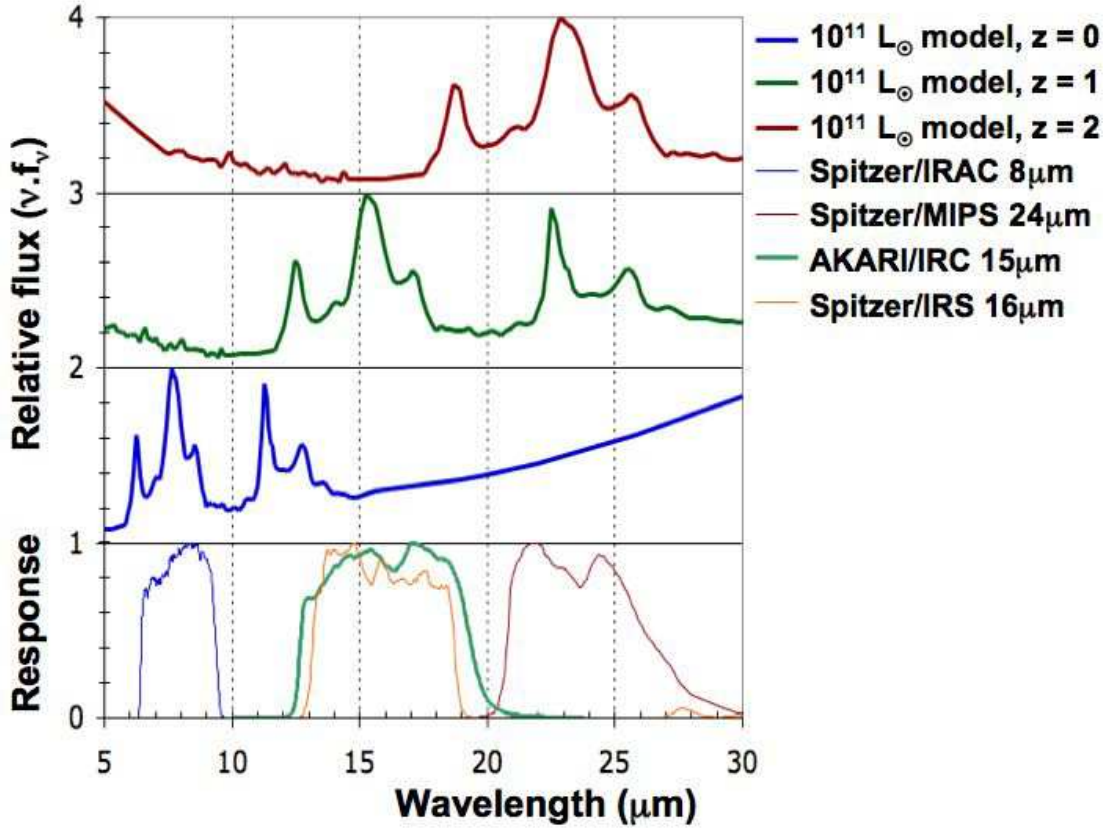}
   \end{center}
  \caption{The three MIR filters and one model (from \cite{dale02}) at $z = 0$, $z = 1$ and $z = 2$ that show what part of the SED is ntegrated in each filter depending of the redshift.}\label{filters}
\end{figure*}

\begin{figure*}
  \begin{center}
   \FigureFile(15cm,15cm){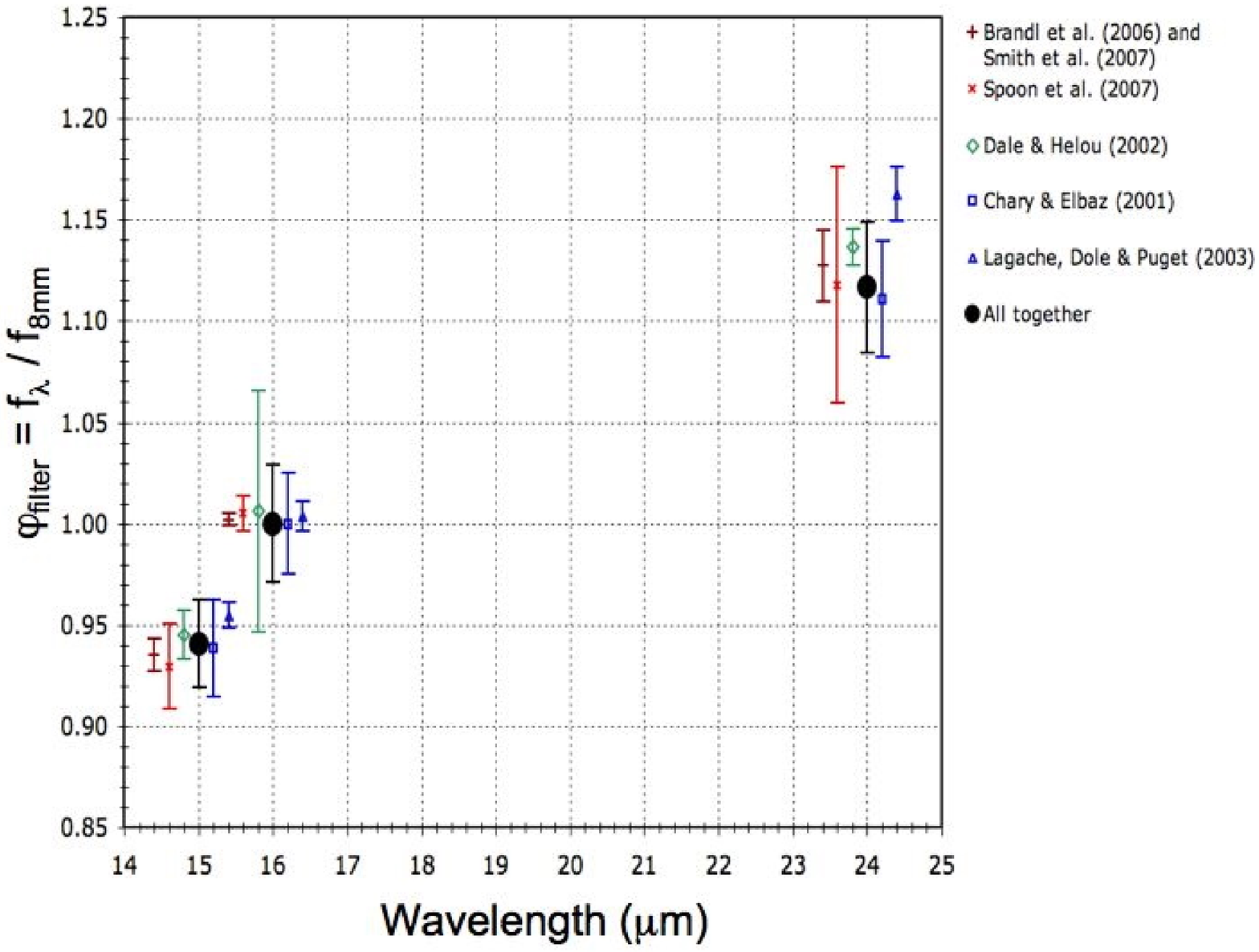}
  \end{center}
  \caption{Corrections $\phi_{filter}$ to apply to the observed MIR photometry for the AKARI/IRC $15 \mu m$ filter and the Spitzer/IRS $16 \mu m$ filter for galaxies at $z = 1$ and for the Spitzer/MIPS $24 \mu m$ filter for galaxies at $z = 2$.}\label{Corrs}
\end{figure*}

Several paper provide a calibration of the rest-frame $8\mu m$ luminosity into a dust luminosity (e.g. \cite{chary01}, \cite{takeuchi05a}, \cite{reddy06}). Given the availability of deep $24 \mu m$ observations, these data have been used to estimate dust luminosities at high redshifts (e.g. \cite{reddy06}). However, if we wish to apply the $z = 0$ conversions to higher redshift galaxies, we must account for corrections because the Spitzer/IRAC $8 \mu m$ filter at $z = 0$ does not sample exactly the same part of the SED than the AKARI $15\mu m$ filter or the Spitzer/IRS $16 \mu m$ filter at $z = 1$ or than the Spitzer/MIPS $24 \mu m$ filter at $z = 2$ (Fig. \ref{filters}). Moreover, these corrections will depend on the SED of the observed galaxies. Corrections are evaluated by convolving several templates and spectra from \citet{chary01}, \citet{dale02}, \citet{lagache04}, \citet{smith07}, \citet{spoon07} and \citet{brandl06}. By using several templates and observed SEDs, we aim at taking into account the variety of observed SEDs. The observed $L_{dust}$ distribution extends from $10^9$ to $10^{11.5} L_{\odot}$, so only models in this luminosity range are used to build Fig. \ref{Corrs}. The corresponding median corrections are listed in Tab. \ref{Tab:Corrs}. We find that the dispersion in the correction factors ($\phi_{filter}$ ) is relatively low (a few percents) for each filter and close to  $\phi_{filter} = 1$, which brings some confidence in using $L_{dust} (8\mu m)$. The correction factors will be used to compare the different estimates of $L_{dust}$ at various redshifts. The redshift range of our objects is $0.9 \le z \le 1.3$. In this range, PAH features can move in and out of the filter. We need to check whether this can influence the results presented in this paper and, if so, what is the size of any effects. To perform this test, we have convolved SEDs from \citet{dale02} and from \citet{chary01} with the AKARI $15 \mu m$ filter. Since all filters will sample the same rest-frame wavelength range, this should address our concern. The values $(s_{z=0.9} - s_{z=1.1}) / s_{z=1.1}$ and $(s_{z=1.3} - s_{z=1.1}) / s_{z=1.1}$ are computed and shown Tab. \ref{Tab:KCorrs}. SEDs from \citet{chary01} provide somewhat larger relative change by about 16\% as compared to the 5\% from  \citet{dale02} SEDs. We can conclude that the redshift range is small enough to keep within the uncertainties even by assuming a single redshift.

\begin{table*}
  \caption{To use rest-frame $8\mu m$ conversions from monochromatic luminosities to $L_{dust}$, we must apply some corrections $\phi_{filter}$ to the photometry. The corrections ($f^{corr}_{\nu} = f^{obs}_{\nu} / \phi_{filter}$) for $z = 1$ for galaxies observed in the AKARI/IRC L15 $15 \mu m$ filter and in the Spitzer/IRS $16\mu m$ filter and for $z = 2$ for galaxies observed in the Spitzer/MIPS $24\mu m$ filter are listed here. }\label{Tab:Corrs}
  \begin{center}
    \begin{tabular}{cccc}
      \hline
Filter               & $z = 1$ AKARI $15\mu m$  & $z = 1$ Spitzer $16\mu m$ & $z = 2$ Spitzer $24\mu m$ \\
$\phi_{filter}$ & $0.94 \pm 0.02$                  & $1.00 \pm 0.03$              & $1.12 \pm 0.03$                  \\
      \hline
    \end{tabular}
  \end{center}
\end{table*}

\begin{table*}
  \caption{We have computed the ratio of the flux densities in the AKARI/IRC L15 $15 \mu m$ filter using \citet{dale02} and from \citet{chary01} $(s_{z=0.9} - s_{z=1.1}) / s_{z=1.1}$ and $(s_{z=1.3} - s_{z=1.1}) / s_{z=1.1}$ to test if the influence of PAH moving with the redshift of the observed galaxies might change the observed flux densities and produce an effect only due to K-corrections.  }\label{Tab:KCorrs}
  \begin{center}
    \begin{tabular}{ccc}
      \hline
SEDs                 &  $(s_{z=0.9} - s_{z=1.1}) / s_{z=1.1}$ & $(s_{z=1.3} - s_{z=1.1}) / s_{z=1.1}$  \\
\citet{dale02}   &  $- 0.092 \pm 0.047$  & $- 0.033 \pm 0.013$   \\
\citet{chary01} & $- 0.161 \pm 0.028$  &   $+ 0.016 \pm 0.035$ \\
      \hline
    \end{tabular}
  \end{center}
\end{table*}

\section{Testing the rest-frame $8 \mu m$ luminosity conversions into total dust luminosity}

Estimating the star formation rate of high redshift galaxies always is an issue, especially when one wishes to estimate the hidden star formation rate or, in other words the dust luminosity. Of course, if FIR data are available as for galaxies detected in IR surveys, the task is more straightforward since we are close to the maximum of the dust emission, but these data are, most often not observed for ultraviolet-selected galaxies like LBGs. Even though $8 \mu m$ luminosities are calibrated into star formation rates, we can wonder if we can apply these calibrations and, if so, what is the associated uncertainty. We directly compare in this paper calibrations for high redshift galaxies which are likely to share common physical characteristics in terms of dust temperature distribution, metallicity, PAH contribution, etc.

\cite{caputi07} performed such a comparison at $z = 2$ on relationships derived by convolving templates with the transmission of the Spitzer MIPS-$24 \mu m$ band. \citet{daddi07}, \citet{papovich07} also performed a similar job at $z = 2$. Here, we use a different approach : we use our LBG selected at $0.9 \le z \le 1.2$ for which we have rest-frame $8\mu m$ data (from AKARI IRC $15 \mu m$-band observations) and rest-frame $12\mu m$ data (from Spitzer MIPS-$24 \mu m$ band observations). These LBGs have $11 \le Log L_{dust} \le 12$ which means that they are Luminous IR galaxies (LIRGs with $11 \le Log L_{dust} \le 12$). A direct comparison of the two estimates allows us to quantify the quality and the uncertainty of the former, after making the hypothesis that the latter provides a good estimate at $z \sim 1$. The conversion is largely dependent on the assumptions made or the templates used (e.g. \cite{lefloch05}). By using directly observed data, we wish to verify what the validity of $L_{dust} (8\mu m)$ is to estimate $L_{dust}$ on actual high redshift galaxies at $z = 1$. This work therefore bears some similarities with \cite{marcillac06} at $z = 1$ but our samples are difference and we add the LBG specificity. 

We define $L_{dust}$ as the total infrared luminosity from $8 \mu m$ to $1000 \mu m$. Another definition uses wavelengths in the range $5 \mu m$ to $1000 \mu m$. \citet{bavouzet08} give the ratio $L_{5 - 1000 \mu m} / L_{8 - 1000 \mu m} = 1.07 \pm 0.04$. In this paper, we prefer using an empirical approach in this paper by using conversions that are mainly based observed datasets and do not rely on (or at least minimize the use of) templates or models. Theefore, to estimate $L_{dust}$ from monochromatic MIR lumininosities, we compute two estimates of $L_{dust}$ at a rest-frame wavelength of $12 \mu m$ : \citet{chary01} and \citet{takeuchi05a}. By using these two formulae, we show that our results are not very dependent on the assumed $12 \mu m$ conversion and we assume that $L_{dust} (12\mu m)$ can be used as a safe reference.

\citet{chary01} provide a formula estimated from a sample of local 300 IRAS Bright Galaxy Sample (\cite{soifer89}). Almost all these galaxies are detected at all IRAS wavelengths from 12 to 100 $\mu m$, which means that they form a far-infrared selected sample dominated by relatively cool (i.e. $T_{dust} \approx 50$K) dust galaxies.

\citet{chary01} relationship corresponds to :

\begin{equation}
$$ L_{dust} = 0.89^{+0.38}_{-0.27} L^{1.094}_{12\mu m}$$ 
\label{eq:ce}
\end{equation}

\citet{takeuchi05a} also provide conversions from several mid- to far-infrared luminosities into total dust luminosities $L_{dust}$. The initial sample built by \citet{takeuchi05a} is extracted from the IRAS PSCz (\cite{saunders00}) that contains 15411 galaxies with redshifts. Then, 1420 galaxies with good quality flux densities at all IRAS wavelengths were down-selected. Star forming quiescent, normal galaxies dominate this sample. However, the authors found that their relationship provide a good estimate for a wide range of spectral energy distributions. The conversion is estimated by comparing monochromatic luminosities with $L_{dust}$ computed using \citet{sanders96} calibration (the four IRAS wavelengths are used) where $L_{dust}$ is computed by fitting a single temperature dust emissivity model (with an emissivity $\epsilon = \nu^{-1}$) to the flux in all four IRAS bands.

\cite{takeuchi05a} conversion (with a dispersion $\sigma = 0.154$) follows the relationship :

\begin{equation}
$$Log L_{dust} = 1.02 + 0.972 Log L_{12\mu m}$$ 
\label{eq:ttt}
\end{equation}

Then, for the conversion from $8 \mu m$ monochromatic luminosities to $L_{dust}$ we made use of \citet{reddy06}, \citet{bavouzet08} and \citet{wu05}.

\citet{reddy06} defined an optically-selected sample with spectroscopic redshifts at $z \sim 2$. For these galaxies, X-ray data are available through stacking and we have indications that the X-ray luminosities are correlated to the total far-infrared ones (e.g. \cite{rosa-gonzalez07}) for galaxies with $8 \le Log L_{dust} [L_{\odot}] \le 12$. X-ray luminosities are proxies for the total infrared luminosities and thy provide a good check for the conversion that is not dependent on templates. To quantify the relationship between $8\mu m$ luminosities and total infrared luminosities, \citet{reddy06} made use of \citet{elbaz02} data for a sample of 149 galaxies that contains ISO data and IRAS $60$ and $100 \mu m$ measurements. 

\cite{reddy06}relationship is : 

\begin{equation}
$$ L_{dust} = (1.24 \pm 0.35) L_{8\mu m}$$ 
\label{eq:reddy}
\end{equation}

\citet{bavouzet08} samples contain galaxies with spectroscopic redshifts and COMBO 17 photometric redshifts observed at $8 \mu m$, $24 \mu m$, $70 \mu m$ and $160 \mu m$ by Spitzer. This sample can be considered as a far-infrared selected sample with redshifts are up to $z = 1$, but most of them have lower redshifts in the range $z \sim 0.2 - 0.3$. They have added 13 stacked data points that are similar to a mid-infrared selection and 3 stacked point from \citet{zheng07} that are representative of an optically-selected sample. \citet{bavouzet08} integrated the spectral energy distributions in rectangles to avoid biases to estimate total dust luminosities.

Finally, the relationship finally defined by \citet{bavouzet08} (with an uncertainty of 37 \%) is : 

\begin{equation}
$$L_{dust} = 450.9 \times (L_{8\mu m}) ^ {0.83}$$.
\label{eq:bavouzet}
\end{equation}

\citet{wu05} exploited the Spitzer First Look Survey imaged by Spitzer IRAC and MIPS and also covered by the Sloan Digital Sky survey (\cite{stoughton02}), which provides $H\alpha$ fluxes for this galaxy sample. One of the advantges of having some spectroscopy is that narrow-lines AGNs can be removed from the sample. All the objects in this sample are star-forming galaxies. The $H\alpha$ fluxes are corrected for dust attenuation computed from the Balmer decrement. A good correlation is observed between the $8 \mu m$ luminosities and the $H\alpha$ luminosities. To deduce the relationship between $\nu . f_{8\mu m}$ and $L_{dust}$, we convert their star formation rate into $Log L_{dust}$ with \citet{kennicutt98} SFR formula.

\citet{wu05} relationship corresponds to :

\begin{equation}
$$ L_{dust} = 0.566 L_{8\mu m}$$
\label{eq:wu}
\end{equation}

\begin{figure*}
  \begin{center}
   \FigureFile(18cm,18cm){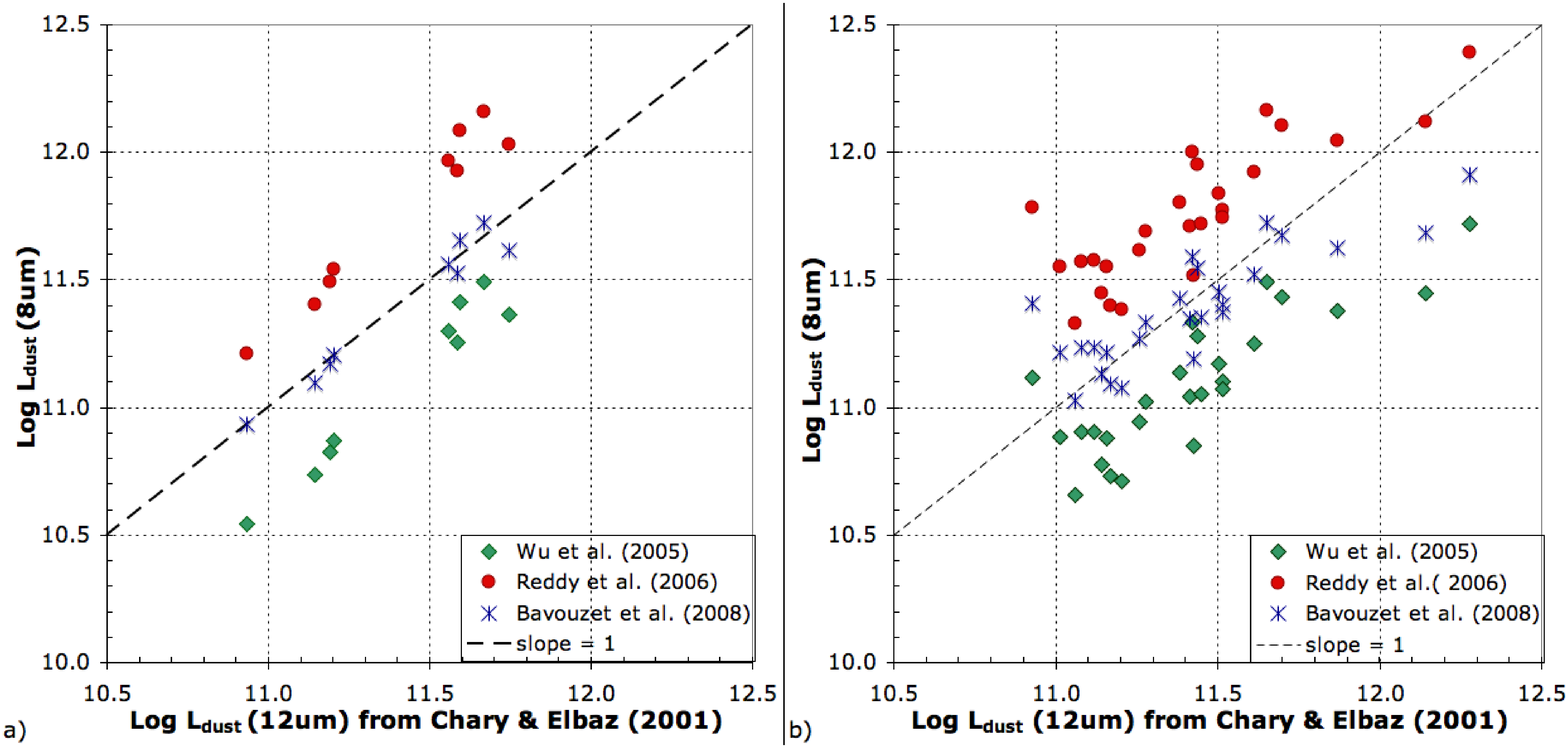}
  \end{center}
\caption{The two panels compare the relative performances of several conversions of $L_{8\mu m}$ into $L_{dust}$ as compared to $L_{dust} (12\mu m)$ one (assumed to be more reliable because of a more dominant dust emission) from \citet{chary01}. Red dots correspond to the comparison of \citet{reddy06} $8\mu m$ conversion with \citet{chary01} $12\mu m$ one. Blue crosses correspond to the comparison of \citet{bavouzet08} $8\mu m$ conversion with \citet{chary01}  $12\mu m$ one. And green diamonds correspond to the comparison of \citet{wu05} $8\mu m$ conversion with \citet{chary01}. The left-hand panel (a) contain the LBG sample while the right-hand panel (b) contain the CDFS AKARI $8 \mu m$ selection. }\label{Ldust_8vs12CE}
\end{figure*}

\begin{figure*}
  \begin{center}
   \FigureFile(18cm,18cm){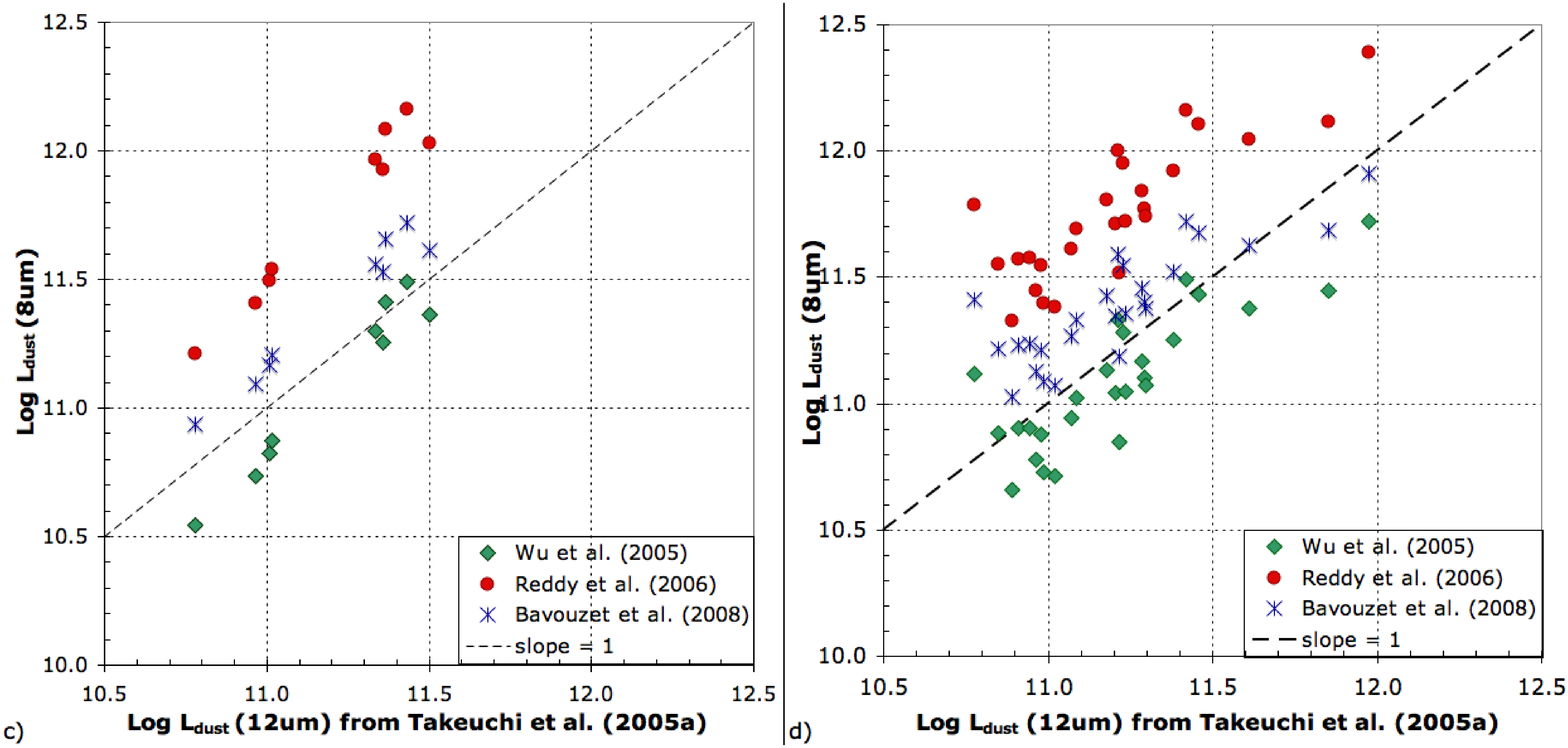}
  \end{center}
\caption{As for Fig. 11 but for \cite{takeuchi05a}.}\label{Ldust_8vs12TTT}
\end{figure*}

The two left panels of Figs. \ref{Ldust_8vs12CE} and \ref{Ldust_8vs12TTT} refer to the LBG sample while the two right panels are related to the CDS AKARI MIR-selected sample.

The first obvious point is that all conversions presented in Figs. \ref{Ldust_8vs12CE} and \ref{Ldust_8vs12TTT} are not fully consistent and we can see differences on the resulting $8 \mu m$ conversions into $L_{dust}$. We clearly observe a stratification of the $8 \mu m$ conversions (as compared to the $12 \mu m$ ones : \cite{chary01} and \cite{takeuchi05a}) from \citet{reddy06} that over estimate $L_{dust}$ both for LBGs and for the AKARI sample to \citet{bavouzet08} conversion that provides globally the best estimate and to \citet{wu05} conversion that slighly under estimates $L_{dust}$. The conversion formula from \citet{wu05} is calibrated from $H\alpha$ data corrected for dust attenuation using the Balmer decrement. Obtaining an under estimated $L_{dust}$ is not unexpected since $H\alpha$ observations might miss some deeply enshrouded emissions.

We see a difference between the LBG and the MIR-selected diagrams. While for the former, the slopes of all the conversions are close to 1 (see Tab. \ref{tab:Ldust_corr}), they are shallower for the MIR selection. It is interesting to notice that the relationship is almost perfect between \citet{chary01} and \citet{bavouzet08}. Both of them provide the same $L_{dust}$ within the uncertainties. On the other hand, the slope is shallower for the MIR-selected sample. The origin of this difference seems to be due to the extremes at low ($L_{dust} \sim 10^{11} L\odot$) and high ($L_{dust} \ge 10^{12} L_\odot$, i.e. ULIRGs) dust luminosities. Taking off high-$L_{dust}$ galaxies in the fit does not change the slope which remains significantly different from 1. To reach similar slopes, we have to also take off the lowest luminosity galaxies from the MIR sample. Part of the difference in the slope can be explained by the fact that LBGs at $L_{dust} \sim 10^{11} L_\odot$ have lower $L_{dust}$ estimated from $8 \mu m$ than the AKARI sample. A detailed analysis and a comparison to models will be performed in the next section devoted to the MIR SEDs and to the interpretation of the $f_{24} / f_{15}$ flux ratio that might provide some explanation for this difference.

\begin{table*}
  \caption{Statistical information on the regressions between the rest-frame $8\mu m$ and $12\mu m$ $L_{dust}$ estimates for a linear fitting $log L_{dust} (12\mu m) = a + log L_{dust} (12\mu m)$. }\label{tab:Ldust_corr}
  \begin{center}
    \begin{tabular}{llccc}
      \hline
Selection & $L_{dust} (8\mu m$) vs. $L_{dust} (12\mu m$) & a & b & Correlation coefficient \\
LBG              & \citet{chary01} vs. \citet{wu05}                & $1.183 \pm 0.095$ & $-2.400 \pm 1.085$ & 0.957\\
LBG              & \citet{chary01} vs. \citet{reddy06}            & $1.183 \pm 0.095$ & $-1.731 \pm 1.085$ & 0.957 \\
LBG              & \citet{chary01} vs. \citet{bavouzet08}      & $0.982 \pm 0.079$ & $0.192 \pm 0.900$ & 0.957 \\
LBG              & \citet{takeuchi05a} vs. \citet{wu05}           & $1.331 \pm 0.107$ & $-3.818 \pm 1.199$ & 0.957 \\
LBG              & \citet{takeuchi05a} vs. \citet{reddy06}       & $1.331 \pm 0.107$ & $-3.149 \pm 1.199$ & 0.957 \\
LBG              & \citet{takeuchi05a} vs. \citet{bavouzet08} & $1.105 \pm 0.089$ & $-0.985 \pm 0.995$ & 0.957 \\
$15\mu m$ & \citet{chary01} vs. \citet{wu05}                & $0.685 \pm 0.093$ & $3.267 \pm 1.057$ & 0.696 \\
$15\mu m$ & \citet{chary01} vs. \citet{reddy06}            & $0.685 \pm 0.093$ & $3.936 \pm 1.057$ & 0.696 \\
$15\mu m$ & \citet{chary01} vs. \citet{bavouzet08}      & $0.569 \pm 0.077$ & $4.896 \pm 0.877$ & 0.696 \\
$15\mu m$ & \citet{takeuchi05a} vs. \citet{wu05}           & $0.771 \pm 0.104$ & $2.445 \pm 1.167$ & 0.696 \\
$15\mu m$ & \citet{takeuchi05a} vs. \citet{reddy06}       & $0.771 \pm 0.104$ & $3.115 \pm 1.167$ & 0.696 \\
$15\mu m$  & \citet{takeuchi05a} vs. \citet{bavouzet08} & $0.640 \pm 0.086$ & $4.214 \pm 0.969$ & 0.696 \\
      \hline
    \end{tabular}
  \end{center}
\end{table*}

\section{The evolution of the $L_{dust} / L_{UV}$ ratio of ultraviolet-selected galaxies with the redshift}

\begin{figure*}
  \begin{center}
   \FigureFile(15cm,15cm){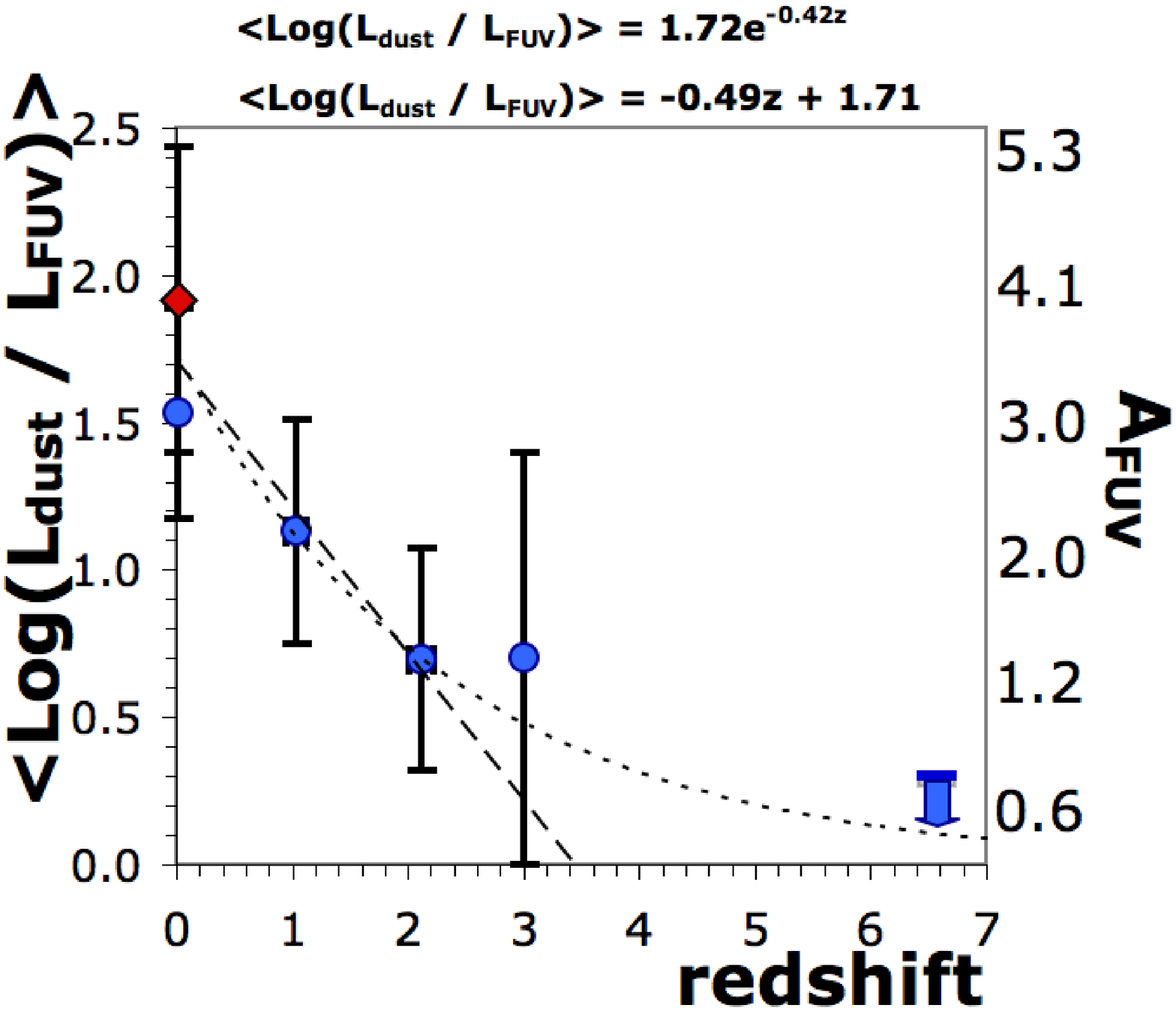}
  \end{center}
\caption{Evolution of $L_{dust}/L_{FUV}$ ratio for UV-selected samples as a function of the redshift from $z \sim 0$ (\cite{buat07}) to $z \sim 1$ (\cite{burgarella07}) and to $z \sim 2$ (\cite{reddy06}) using the new conversion presented in the text). A point at $z \sim 3$ from \citet{adelberger00} and an upper limit from a galaxy at $z = 6.56$ (\cite{boone07}). Both an exponentially decaying law (dotted line) and a linear regression (dashed line) are a fair representation of the decline of the FUV dust attenuation in this redshift range.At $z = 0$, the red point corresponds to an IR-selected sample and the blue one to an extrapolation from the UV-selected sample.}\label{LdustLuv}
\end{figure*}

\citet{burgarella07} and \citet{reddy06} found a decrease of the $L_{dust} / L_{UV}$ ratio of ultraviolet-selected galaxies with the redshift from $z = 0$ to $z = 2$. However, since we found in the previous section that \citet{reddy06} over estimated $L_{dust}$, we must revisit the evolution of $L_{dust} / L_{UV}$ assuming another conversion. Note that with the updated calibration, the median and mean $L_{dust}$ for Reddy et al.'s sample are respectively $1.4~10^{11} L_\odot$ and $1.4~10^{11} \pm 4.3~10^{11} L_\odot$. We also find in the previous section the conversion given by \citet{wu05} provides under estimated $L_{dust}$, probably because of some missing emission too deeply enshrouded in dust. We will therefore use \citet{bavouzet08} conversion instead of \citet{reddy06} in the following to compute a new estimate of $L_{dust} / L_{UV}$ at $z = 2$. This new conversion means that $Log~L_{dust}$ decreases by $-0.37 \pm 0.05$ for ultraviolet-selected galaxies in the luminosity range corresponding to LIRGs. The $z = 1$ $L_{dust} / L_{UV}$ was computed by \citet{burgarella07} by using \citet{chary01} conversion and we are, therefore, in the optimum case where both estimates from $8\mu m$ and $12 \mu m$ $L_{dust}$ are fully consistent. Note that replacing \citet{chary01} conversion by \citet{takeuchi05a} one would change $L_{dust}$ estimated from the rest-frame $12 \mu m$ luminosityes by $-0.2 dex$. Fig. \ref{LdustLuv} updates the one presented in \citet{burgarella07}. The change applied to \citet{reddy06} at $z = 2$ amplifies the decrease with the redshift. We have added one point at $z \sim 3$ from \citet{adelberger00} estimated from their Fig.~11 at $L_{dust} / L_{UV} = 0.7 \pm 0.7$. The dust luminosity of their sample is about $10^{10} - 2.10^{12} L_\odot$. Although the $L_{dust} / L_{UV}$ was estimated from the UV slope $\beta$ (\cite{meurer99}), \citet{burgarella07} suggests that it provides a rough order of magnitude that is a fair estimate for galaxies in an intermediate $L_{dust}$ range (but with a large dispersion). Together with the addition of a point at $z = 6.56$ from the upper limit on the gravitational lens HCM 61 studied by \citet{boone07}. It is important to notice that the dust luminosity of this very high redshift object is $L_{dust} < 2.1 10^{11} L_\odot $. So, all the luminosities lie in the same range - mainly LIRGs. Fig. \ref{LdustLuv} seems to confirm the global decreasing trend with the redshift for these galaxies with dust luminosities lower than ULIRGs. If we follow the authors who suggest that the object is close to a detection, an exponential law is perhaps favored.

\section{Testing the MIR Spectral Energy Distributions with the $f_{24} / f_{15}$ flux ratio}
\begin{figure*}
  \begin{center}
   \FigureFile(21cm,15cm){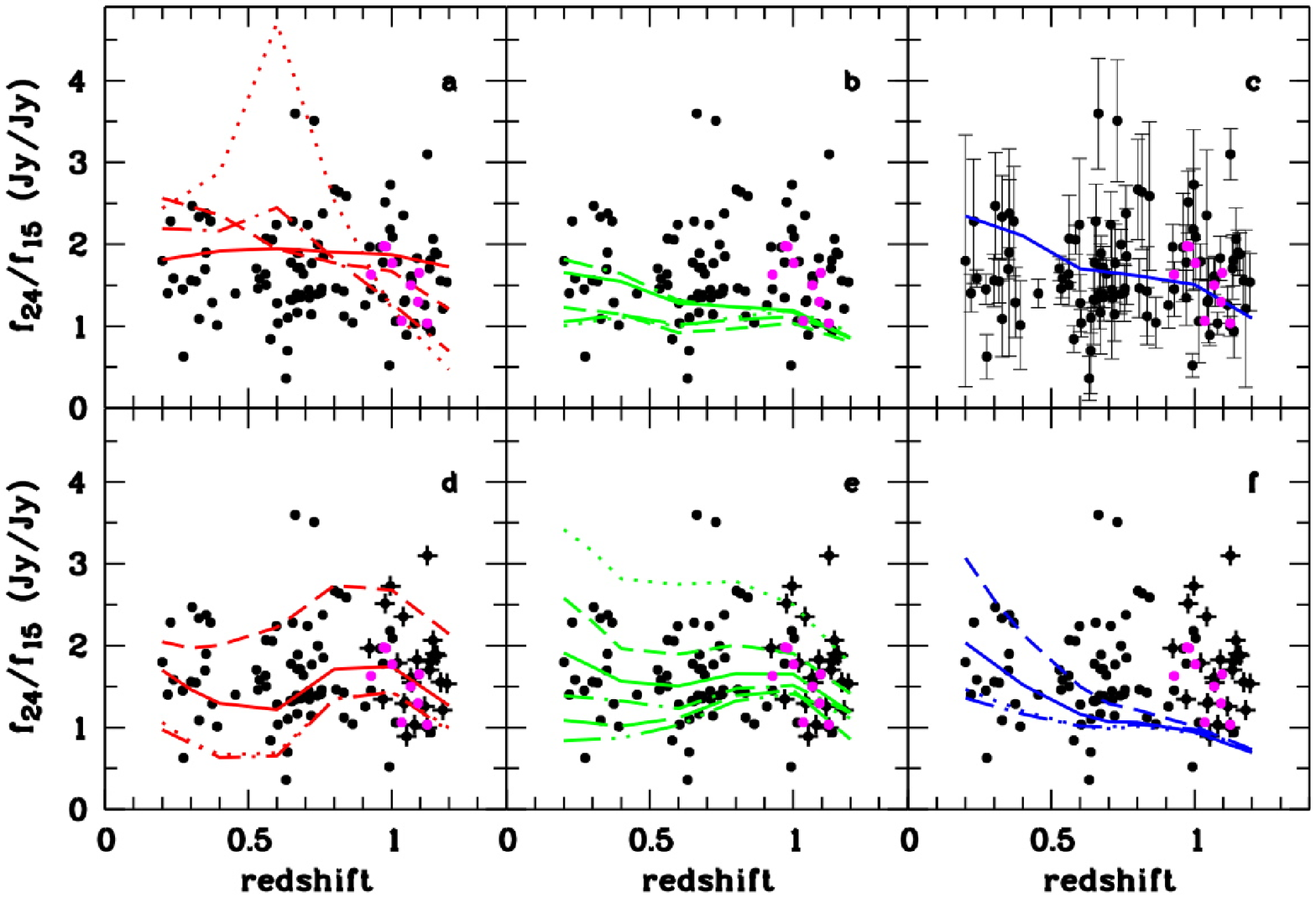}
   \end{center}
  \caption{$f_{24} / f_{15}$ ratio as a function of redshift. Observed ratios are plotted as dots, in panels d,e and f crosses are used for galaxies with L$_{dust}> 10^{11}$ L$_\odot$ and z$>$0.8, error bars on the measurements are reported in panel c only to avoid crowded plots. The templates are overplotted as lines in the different panels:{\bf a} IRS templates representative of three classes of objects from \citet{spoon07}: hot dust dominated (1A, solid),PAH dominated (1C (dash), 2C (dot-dash), silicate absorption dominated (3B, dotted); {\bf b}, IRS templates of star-forming galaxies from \citet{smith07}: luminosity averaged template (solid) and templates 1 (long dash), 2 (dotted), 3 (dash) and 4 (dot-dash); {\bf c}, IRS template for a typical starburst galaxy from \citet{brandl06}; {\bf d}, CE templates for  $10^9$ (dot-dash), $10^{10}$ (dotted), $10^{11}$ (solid) and  $10^{12}$ $L\odot$ (dash);{\bf e} DH templates with $f_{60}/f_{100}$=0 (dotted),-0.1 (dash),-0.2 (solid),-0.3 (dot-dash),-0.4 (lond dash),-0.5 (dash-dot-dot); {\bf f}: LDP templates, same luminosities and lines as for CE templates (panel d)}\label{zf2415}
\end{figure*}

The  bandpasses of the AKARI/IRC and Spitzer/MIPS filters are larger than the PAH features and do not allow us to perform a detailed analysis of the evolution with redshift of the MIR SEDs of galaxies. Nevertheles, we can check the consistency of the available libraries with our set of data since from z=0 to z=1.2 the IRC 15 $\mu$m and MIPS 24 $\mu$m filters encounter the PAH features (cf Fig. \ref{filters}). Such an analysis may also help us to understand the different behaviors of the LBG and MIR-selected samples in the previous sections where the various calibrations of the 8 $\mu m$ flux in bolometric flux are discussed.

Following \citet{marcillac06} and \citet{teplitz05} we  compare the observed 24 to 15 $\mu m$ flux ratio  $f_{24} / f_{15}$ with the expected ones calculated to the templates presented in the previous section. The predicted fluxes are calculated by integrating the templates within the filter bandpasses.   

\subsection{Variation of $f_{24} / f_{15}$ with z}

In Fig. \ref{zf2415} we compare the observed variation of $f_{24} / f_{15}$ with z for both samples ($15\mu m$ MIR selection and LBGs) with the predictions of the differents models presented in  section 5. The ratio exhibits a large dispersion without any significant trend. Error bars are overplotted in the sub-panel (c).  The templates used for the comparison are chosen to sample a large range of luminosity and properties. From the CE and LDP libraries, we consider four templates  corresponding to  $10^9$, $10^{10}$, $10^{11}$ and  $10^{12}$ $L\odot$. Six values of $f_{60}/f_{100}$ (0,-0.1,-0.2,-0.3,-0.4,-0.5) are selected from the DH library in order to cover  a large range of $L_{dust}$ values (from $5~10^8$ to $4~10^{12} L_\odot$ according to the calibration of \citet{marcillac06}, $f_{60}/f_{100}$ increasing monotically with $L_{dust}$) \\

At z$>0.5$ the DH and CE libraries are able to cover the range of observed values although some galaxies at z$\sim$1 exhibit a $f_{24} / f_{15}$ ratio lower than that predicted by the models. A substantial fraction of the galaxies detected at 15 and 24 $\mu$m  are LIRGs, this is illustrated in  Fig. \ref{zf2415} for z$>$0.9 galaxies (see the next section for the calculation of L$_{dust}$). If we restrict the CE templates to L$_{dust} > 10^{11}$ L$_\odot$,  galaxies with  $f_{24} / f_{15} < \sim 1.5$ and z$>$0.9 are not well represented by them : templates  with a lower L$_{dust}$ are needed to reach such low values of $f_{24} / f_{15}$ which are observed for $\sim 30\%$ of the MIR-selected galaxies with $z > 0.9$, this issue was  already underlined by \citet{teplitz05} and will be re-discussed in the next section. The LDP templates predict lower observed  ratios which can fit the lowest observed values  but are unable to reproduce the bulk of the sample at z$>$0.5.

Galaxies at z$\sim$1 with a large $f_{24} / f_{15}$ ratio can be explained either by a starburst template with a strong continuum together with PAHs (CE template with $10^{12}$ L$_\odot$, DH templates with $\log(f_{60} / f_{100})$=0,-0,1 also corresponding to galaxies with L$_{dust}\sim 10^{12}$ L$_\odot$)  or by a SED without PAHs and a power-law spectrum. This fitting degeneracy is also underlined by \citet{marcillac06} and \citet{elbaz05}

The comparison with the IRS templates implies that the spectra of normal star forming galaxies from \citet{smith07} do not sample the whole set of data and are only able to reproduce the lowest observed values whereas the starburst template of \citet{brandl06} goes through the data. The different templates of \citet{spoon07} exhibit  a larger dispersion for the evolution of $f_{24} / f_{15}$ with z as expected since they sample extreme classes of objects. Templates dominated by the silicate absorption do not reproduce at all the general trend of the data: such galaxies are rarely found  in a MIR selection up to z=1.

The LBG sample exhibits less dispersed and lower values in average of $f_{24} / f_{15}$  than the MIR-selected sample. Their behaviour is consistent with that predicted for galaxies fainter than  $10^{12}$ L$_\odot$ (CE templates) or with $\log(f_{60} / f_{100})<-0.1$ (DH templates). The starburst template of  \citet{brandl06} reproduces well the mean behaviour of the LBGs

To summarize, starburst templates and  SED libraries for bright IR galaxies ($\ge \sim 10^{11}$ L$_\odot$) are globally consistent with most of the  $f_{24} / f_{15}$ ratios observed for MIR-selected at z$>$0.5. About $\sim 30\%$ of these MIR-selected galaxies at z$>$0.9 exhibit  values of $f_{24} / f_{15}$ lower than those for starburst and IR bright galaxies and more consistent with the ratios found  for normal star forming galaxies and templates with a moderate luminosity, as well as with the LDP library. Nevertheless, most of these galaxies are LIRGs. LBGs at z$\sim$1 exhibit  intermediate $f_{24} / f_{15}$ ratios not as large as those predicted for bright IR galaxies and galaxies with a strong continuum. They are more typical of starburst galaxies as expected for LBGs.

For z$<$0.5 galaxies, IRS spectra of normal star-forming galaxies \citep{smith07} or SED templates for moderately bright galaxies (CE, DH, LDP) reproduce the data quite well. Strong starbursts or continuum dominated distributions \citep{brandl06,spoon07} are not consistent with most data.

\subsection{ Variation of $f_{24} / f_{15}$ with $L_{dust}$}

\begin{figure*}
  \begin{center}
   \FigureFile(24cm,18cm){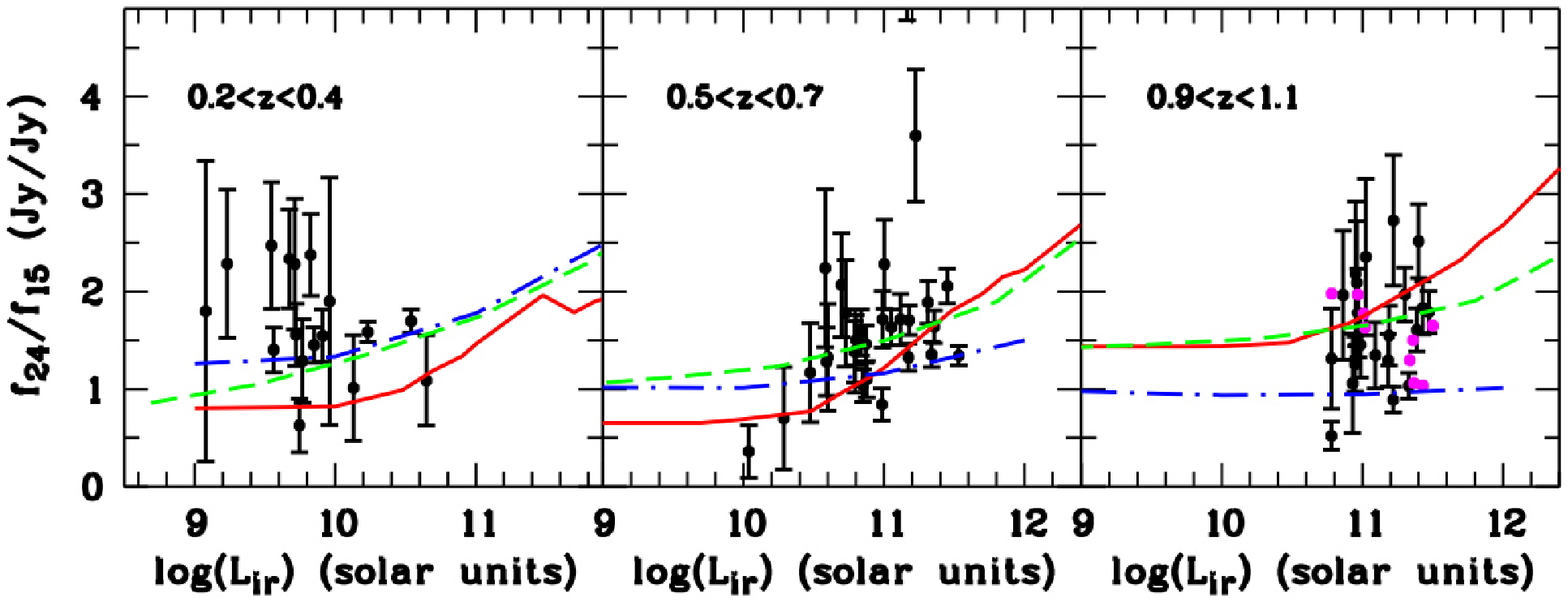}
   \end{center}
  \caption{$f_{24} / f_{15}$ ratio as a function of L$_{dust}$ for three differents redshift bins. The black dots and error bars are for the MIR-selected sample. The LBGs are represented with cyan dots in the third panel, error bars are not represented but are similar to those for the IR selected sources. The predictions of the models are overplotted as lines: CE templates (red solid), DH templates (green dash) and LDP templates (blue dot-dash)}\label{Lir-f2415}
\end{figure*}

As underlined above,  the various libraries of templates can reproduce the full dynamics of the $f_{24} / f_{15}$ ratio but not necessarily with the good luminosity. Therefore, we have to check their consistency   in terms of luminosity. Indeed, the CE and LDP templates vary as a function of L$_{IR}$ and  the parameter characterizing the DH templates, $f_{60} /f_{100}$, can be roughly related to an IR luminosity \citep{marcillac06}.

As discussed in Sect. 5, the total IR luminosities L$_{dust}$ of the observed galaxies are estimated using empirical relations between L$_{dust}$ and a monochromatic mid-IR luminosity  calibrated on local samples from \cite{takeuchi05a}. Practically, calibrations are available for rest-frame data at 12 and 15 $\mu$m. Therefore we will restrict our analysis to  three redshift bins: $0.2<z<0.4$ (fluxes measured with the IRC 15 $\mu$m filter at $z = 0.3$ correspond to 12 $\mu$m rest-frame since the isophotal wavelength of the L15 filter is 15.6 $\mu$m), $0.5<z<0.7$ and $0.9<z<1.1$ (fluxes measured at $z = 0.6$ and $z = 1$ with the MIPS 24 $\mu$m filter (isophotal wavelength $23.7 \mu$m) correspond respectively to $15 \mu$m and $12 \mu$m rest-frame). Using the relations of \citet{chary01} will give only slightly different luminosities which will not have any influence on the following analysis.

In Fig. \ref{Lir-f2415} are plotted the observed ratios together with the predictions of the CE, DH and LDP models for the three redshift bins. As expected we select brighter galaxies when z increases. At low z ($0.2<z<0.4$) the predictions of the models are consistent with a large fraction of the data but 5 out of the 17 galaxies observed in this redshift bin exhibit ratios larger than those predicted (even when error bars are accounted for). Such large values are obtained for IRS spectra of strong starburst galaxies as can be seen in Fig.\ref{zf2415} (panels a and c). At this redshift the observed $f_{24} / f_{15}$ corresponds to  $f_{18} / f_{12}$ in the rest-frame of the galaxies: these galaxies are likely to exhibit a strong continuum implying a high flux at 18 $\mu$m. For the medium redshift bin ($0.5 \le z \le 0.7$) the models are found globally consistent with the data given the error bars on the observed ratios. One galaxy is found very discrepant, its ratio is only consistent with a template dominate by a strong silicate absorption (Fig.\ref{zf2415} (panel a)). At high z ($0.9<z<1.1$), both MIR-selected galaxies and LBGs are considered since 9 LBGs are detected in our AKARI image in the redshift bin $0.9 \le z \le 1.1$. The CE, DH and LDP models bracket the observations. Given the large dispersions of the observed values the CE and DH templates can be considered as globally consistent with the data confirming the global reliability of the CE relations to estimate $L_{dust}$ from the monochromatic luminosity at 8 and 12 $\mu$m rest-frame (see Sect. 6). Half of the sample (9/21) correspond to  $f_{24} / f_{15}<1.5$ whereas the CE and DH models predict higher ratios ($f_{24} / f_{15}>1.5$) for the observed range of luminosity; the LDP models are consistent with the lowest observed ratios. At z$\sim$1 the observed $f_{24} / f_{15}$ corresponds to $f_{12} / f_{8}$, as underlined by \citet{elbaz05} such low values of this ratio can only be reproduced with a strong PAH emission feature at 7 $\mu$m. To explain the large variations of this ratio one might invoke the relative strengths of the PAH features or the contribution of the underlying continuum.  On the one hand, the galaxies with a low $f_{24} / f_{15}$ ($f_{24} / f_{15}<\sim 1.5$) may have a weaker 12 $\mu$m PAH emission feature  as compared to the PAH feature at $7\mu m$. Indeed \citet{lagache04} modified the MIR templates in this way and their templates can explain the most extreme values of our sample. On the other hand, the CE and DH models for bright galaxies (as those observed at z$\sim$1) exhibit  a high continuum which contributes to  large  predicted values of $f_{24} / f_{15}$ without a strong variation of the PAH feature at 12 $\mu$m; in Fig. \ref {Lir-f2415} the CE and DH templates of lower luminosity exhibit a $f_{24} / f_{15}$ more consistent with the data.  In the same way, \citet{smith07} templates have a rather low continuum and fit the observed low ratios (Fig.\ref{zf2415} (panel b)). \citet{teplitz05} also found that a ULIRG spectrum is not consistent with the variation of the 24 and 16 $\mu$m fluxes observed with Spitzer up to z$\sim$1.5 and that local star-forming or starburst galaxies are  more representative of  the high z  galaxies detected by Spitzer. At z$\sim$0.7 \citet{zheng07} reached a similar conclusion that luminous IR galaxies have relatively low temperature, lower than that of local galaxies of same luminosity  which implies a low contribution of the continuum.

The LBGs at $z = 1$ also exhibit values of the $f_{24} / f_{15}$ ratio characteristic of the presence of PAH features in their spectra, they appear less dispersed than the IR selected galaxies as already underlined. The CE and DH templates reproduce rather well the mean values confirming the good agreement found in Sect. 5 when $L_{dust}$ was calculated using the CE calibrations at $12/mu m$ (rest-frame). Although the sub-sample considered here is small (9 objects) it seems that  $f_{24} / f_{15}$ is found higher in LBGs with $L_{dust}<10^{11} L\odot$ than for LBGs with $L_{dust}>10^{11} L\odot$, this trend is not seen within the MIR-selected sample and most of the MIR-selected galaxies with $L_{dust} < 10^{11}$ have a $f_{24} / f_{15}$ ratio lower that those of LBGs. This effect may explain the lower $L_{dust} (8\mu m)$ estimated for LBGs than for the MIR-selected galaxies for a given $L_{dust} (12\mu m$ in Figs. \ref{Ldust_8vs12CE} and  \ref{Ldust_8vs12TTT} (with $L_{dust} < 10^{11} L_{\odot}$. This different behaviour may be at the origin of the slight differences found in the Sect. 6 when the various calibrations of $L_{dust}$ were compared for both samples. 

\section{Conclusion}
This paper is based on the observations of the Chandra Deep Field South carried out at $15 \mu m$ with AKARI/IRC. We have applied the standard pipeline to process the frame and performed some hotometry to define two galaxies samples with MIR flux densities at $\lambda = 15 \mu m$. A first sample is selected at $15 \mu m$ directly (MIR-selected sample) and a second one is based on the mesurement of the $15\mu m$ flux density of an initial sample of 420 LBGs at $z \sim 1$ built by \cite{burgarella07} with available $24 \mu m$ flux densities from Spitzer/MIPS.

The main results of this work are : 1) the number counts evaluated from the mid infrared-selected are consistent with previous works from deep imaging at $16 \mu m$ by \citet{teplitz05} and \citet{wada07}.  We observe a bump around $200\mu Jy$ that might be shifted to fainter fluxes with respect to other works because of Cosmic variance. By dividing the differential counts into several sub-samples at different redshifts, we find that the bump is mainly due to galaxies at $z > 0.4$  and at $z > 0.8$ for the fainter part of the bump. 2) We have compared the dust luminosities estimated from several estimators of the monochromatic MIR luminosities at a rest-frame wavelength of $8 \mu m$ with $L_{dust}$ estimated from MIPS data at a rest-frame wavelength of $12 \mu m$ assumed to be the reference. We find some discrepancies between the estimates. \citet{bavouzet08} and, to some extent, \citet{wu05} seems to provide dust luminosities consistent with the two $L_{dust}$ estimates from \citet{chary01} and \citet{takeuchi05a}. 3) A comparison of the $f_{24} / f_{15}$ ratio ratio with several templates and observed SEDs is performed that suggest a global consistency of the observed ratios albeit there might be an issue with the luminosity dependence of this ratio. 4) By making use of the best conversion of the MIR luminosity at $z = 2$, we have revisited the evolution of the $L_{dust} / L_{UV}$ ratio with the redshift and we find a more drastic decrease of ths raio at $z \sim 2$ that previously thought by using \cite{reddy06} conversion.  

\section{Acknowledgements}
This work is based on observations with AKARI, a JAXA project with the participation of ESA and the help of CNES. T.T.T. has been supported by Program for Improvement of Research Environment for Young Researchers from Special Coordination Funds for Promoting Science and Technology, and the Grant-in-Aid for the Scientific Research Fund (20740105) commissioned by the Ministry of Education, Culture, Sports, Science and Technology (MEXT) of Japan. A part of this work has been done during the stay of D.B and V.B in Japan which was supported by Program for Improvement of Research Environment for Young Researchers from Special Coordination Funds for Promoting Science and Technology (MEXT). This paper is the result of a collaboration within the D-SIGALE project.

\end{document}